\documentclass[aps,prd,preprint,superscriptaddress,preprintnumbers,nofootinbib]{revtex4-1}
\pdfoutput=1

\usepackage{graphics}

\usepackage[dvips]{graphicx}
\usepackage{mathrsfs}
\usepackage{amssymb}
\usepackage{amsmath}
\usepackage{verbatim}
\usepackage{float}
\usepackage{slashed}
\usepackage{bbm}
\usepackage[dvips,letterpaper,text={6.5in,9in}]{geometry}

\usepackage[english]{babel}
\usepackage{tabu, multirow,color,dcolumn,bm,enumerate}

\usepackage{graphicx,bm}

\newcommand{\yc}{y_{\chi}}

\newcommand{\bea}{\begin{eqnarray}}
\newcommand{\eea}{\end{eqnarray}}
\newcommand{\bal}{\begin{align}}
\newcommand{\eal}{\end{align}}

\newcommand{\beq}{\begin{equation}}
\newcommand{\eeq}{\end{equation}}

\newcommand{\tev}{\mathrm{TeV}}

\newcommand{\cL}{\mathcal{L}}

\newcommand{\Z}{\mathbb{Z}}

\newcommand{\DM}{\text{DM}}
\newcommand{\D}{\text{D}}
\newcommand{\R}{{_R}}
\newcommand{\X}{{_X}}

\newcommand{\SU}{SU(2)_{_R}}

\definecolor{red1}{cmyk}{0,1,1,0.3}

\usepackage{outlines}
\usepackage{enumitem}
\setenumerate[1]{label=\Roman*.}
\setenumerate[2]{label=\Alph*.}
\setenumerate[3]{label=\roman*.}
\setenumerate[4]{label=\alph*.}

\usepackage{tikz}
\usetikzlibrary{decorations.pathmorphing,decorations.markings}
\tikzset{
photon/.style={decorate, decoration={snake,amplitude=4pt, segment length=7pt}, draw=black},
particle/.style={draw=black, postaction={decorate}, decoration={markings,mark=at position .5 with {\arrow[draw=black]{>}}}},
antiparticle/.style={draw=black, postaction={decorate}, decoration={markings,mark=at position .5 with {\arrow[draw=black]{<}}}},
gluon/.style={decorate, draw=black, decoration={coil,amplitude=3pt, segment length=4pt}},
Higgs/.style={draw=black,dashed,thick },
arrow/.style={draw=black, very thick, postaction={decorate}, decoration={markings,mark=at position 1 with {\arrow[draw=black]{>}}}}
}

\usepackage{latexsym}
\usepackage{subcaption}

\usepackage{outlines}
\usepackage{enumitem}
\setenumerate[1]{label=\Roman*.}
\setenumerate[2]{label=\Alph*.}
\setenumerate[3]{label=\roman*.}
\setenumerate[4]{label=\alph*.}

\definecolor{darklightsabergreen}{rgb}{0.0, .49, 0.06}

\begin{document} 

\title{
{\small\begin{flushright}
\end{flushright}}
\vspace{0.5cm}
 Multi-Component Dark Matter in a Non-Abelian Dark Sector}


\author{Fatemeh Elahi\footnote{felahi@ipm.ir}}
\affiliation{School of Particles and Accelerators, Institute for Research in Fundamental Sciences (IPM) P.O. Box
19395-5531, Tehran, Iran}
\author{Sara Khatibi\footnote{sara.khatibi@ut.ac.ir}}
\affiliation{ Department of Physics, University of Tehran, North Karegar Ave. Tehran 14395-547, Iran}
\affiliation{School of Particles and Accelerators, Institute for Research in Fundamental Sciences (IPM) P.O. Box
19395-5531, Tehran, Iran}

\begin{abstract}
\vspace*{0.5cm}

{ In this paper, we explore a dark sector scenario with a gauged $SU(2)_\R$ and a global $U(1)_\X \times \Z_2$, where the continuous symmetries are spontaneously broken to a global $U(1)_\D$. We show that in various regions of the parameter space we can have two, or three dark matter candidates, where these dark matter particles are either a Dirac fermion, a dark gauge boson, or a complex scalar. The phenomenological implications of this scenario are vast and interesting. We identify the parameter space that is still viable after taking into account the constraints from various experiments. We, also, discuss how this scenario can explain the recent observation by DAMPE in the electron-positron spectrum. Furthermore, we comment on the neutrino mass generation through non-renormalizable interactions between the standard model and the dark sector. 
} 
\end{abstract}

\maketitle

\section{Introduction}
\label{intro}

 Despite the numerous successes of the Standard Model (SM) in describing the observed phenomena, there are still intriguing questions that wait to be answered. Arguably, the most important one among them is the nature and origin of dark matter (DM). For some decades, the leading theory was a single component thermal relic with weak size couplings and mass, commonly known as Weakly Interacting Massive Particle. With the advancement of experiments, however, most of the parameter space of a single-component thermal relic has been excluded. Therefore, we are compelled to examine more complex structures of dark sectors. Among the proposed scenarios, multi-component dark matter (MCDM) has attracted a lot of attentions~\cite{Dienes:2011ja,Dienes:2011sa,Bian:2013wna,Duda:2001ae,Duda:2002hf,Profumo:2009tb,Gao:2010pg,Feldman:2010wy,Baer:2011hx,Aoki:2012ub,Chialva:2012rq,Bhattacharya:2013hva,Esch:2014jpa,Bian:2014cja,YaserAyazi:2018lrv,Ahmed:2017dbb,DuttaBanik:2016jzv,DiFranzo:2016uzc,Dienes:2013xff,Biswas:2015sva,Herrero-Garcia:2018qnz,Karam:2016rsz,Bhattacharya:2018cgx,Bhattacharya:2016ysw,Dev:2016qeb,Khlopov:1995pa,Bhattacharya:2017fid,Huang:2015wts}. In these scenarios, the total relic abundance of dark matter is due to the existence of multiple  dark matter species. Given the rather complex structure of the SM, it should not be surprising if the dark sector has multiple species as well, but to further motivate MCDM scenarios, the extra degrees of freedom in the dark sector are usually employed to explain some other shortcomings of the SM. 

The most common approach in MCDM models is assuming one or multiple symmetries in the dark sector. MCDM models with a gauged $U(1)$ extension or a conserved non-abelian gauge symmetries have already received some attention~\cite{Ahmed:2017dbb,DuttaBanik:2016jzv,Karam:2016rsz,Bhattacharya:2018cgx,Bhattacharya:2016ysw,Davoudiasl:2013jma,Barman:2018esi,Gross:2015cwa,Yamanaka:2015tba,Dev:2016xcp}. In this paper, we focus on a gauged $\SU$ times a global $U(1)_\X $ that are spontaneously broken to a global $ U(1)_D$, once a scalar $\phi$ -- a doublet of $\SU$ with a non-zero charge under $U(1)_\X$-- acquires a vacuum expectation value (vev). Due to this breaking, we have three massive gauge bosons ($W_\R^\mu$). We further assume that dark sector respects a $\Z_2$ symmetry that stays conserved after the spontaneous symmetry breaking. This $\Z_2$ symmetry becomes crucial in making sure we have multiple DM species in various regions of the parameter space. To extend the dynamics of the dark sector, we assume there exists another scalar ($\eta$), and two Dirac Fermions ($\chi_1$ and $\chi_2$), some of which have the potential to be a dark matter candidate. 

The communication of dark sector with the SM content can occur through various means (e.g., kinetic mixing, scalar portal, etc). The kinetic mixing of non-abelian symmetries with any of the SM gauge symmetries is usually non-renormalizable, leading to small interaction between the particles in the two sectors. Therefore, we mainly focus on the scalar portal induced by $\phi$ and the SM Higgs acquiring vevs. This is in many ways similar to a simple Higgs portal model; however, it has some extra advantages that are listed below:
 
 \begin{itemize}
 \item Large self-interactions between some of the DM candidates: Even though collision-less cold dark matter is successful in describing large scale structures~\cite{Blumenthal:1984bp}, it faces some difficulty describing small scale structures. N-body simulations have shown that Self-Interacting DM can alleviate the small scale structure problems~\cite{Tulin:2017ara,Balducci:2018dms}. On the other hand, from direct detection experiments, we are led to believe that DM has negligible interactions with nucleons~\cite{Messina:2018fmz}. Therefore, the dark sector could have a non-trivial structure, where it can allow strong self-interaction, while the portal between the dark sector and SM is rather weak. This is easily achieved in our model. 
 
 \item The extra bosonic degrees of freedom can be used to alleviate the Higgs Hierarchy problem~\cite{Bian:2013wna,Chakraborty:2012rb,Grzadkowski:2009mj,Karahan:2014ola,Antipin:2013exa,Craig:2013xia,Farina:2013ssa}, rescue the vacuum instability~\cite{Bian:2013wna,Gonderinger:2009jp,Drozd:2011aa,Baek:2012se,Gabrielli:2013hma,Hambye:2013sna} allow strong first order phase transition, which is needed to prevent baryonic asymmetry from washing out after its generation ~\cite{Noble:2007kk,Damgaard:2013kva,Profumo:2014opa}. 
 
 \item  Recently, the DArk Matter Particle Explorer (DAMPE) collaboration released their new measurement of the electron-positron flux in the energy range 25 GeV to 4.6 TeV ~\cite{Ambrosi:2017wek}. The results show a sharp peak above the background around 1.4~$\tev$. The sharpness of the peak suggests that DM from a nearby source is annihilating to $e^+ e^-$~\cite{Yuan:2017ysv,Fan:2017sor,Duan:2017pkq,Gu:2017gle,Cao:2017ydw,Liu:2017rgs,Tang:2017lfb,Chao:2017yjg,Gu:2017bdw,Duan:2017qwj,Jin:2017qcv,Niu:2017hqe,Li:2017tmd,Gu:2017lir,Nomura:2017ohi,Ghorbani:2017cey,Yang:2017cjm,Ding:2017jdr,Liu:2017rgs,Okada:2017pgr,Yao:2018ewe,Beck:2018hau,Wang:2018pcc,Balducci:2018dms,Cao:2017sju,Cao:2017rjr,Huang:2015wts}. Assuming that the excess is indeed due to the interaction of DM with electrons, to achieve the height of the resonance, the annihilation cross section needs to be much larger than that of the canonical single component thermal relic. To enhance the cross section of dark matter candidates with electrons, we also charge right-handed electron under $SU(2)_\R \times U(1)_\X$. Even though the main motivation for distinguishing right-handed electron is the results of the DAMPE experiment, the annihilation of dark matter candidates to a pair of electron-positron plays a crucial role in setting the relic abundance. 
   
 \item Neutrino mass generation: Another important observation that cannot be justified within the context of the SM is the mass of neutrinos. In the most minimalistic scenario, we can use the Weinberg operator: $(LH)^2/\Lambda$~\cite{Weinberg:1979sa}, where $\Lambda$ refers to the mass of a heavy Majorana Fermion. A simple calculation reveals that $\Lambda$ has to be bigger than $10^{14} \ \text{GeV}$~\cite{Mohapatra:1979ia}, which is larger than the Landau pole, and in the regime where we cannot trust the SM framework. With a more complex dark sector, we can connect the mass of neutrinos to some of the degrees of freedom in the dark sector. We still use non-renormalizable operators to get a neutrino mass; however, we find a smaller value for the cut-off scale. 
 \end{itemize} 
 In the following section, we explain the model in greater details and introduce the dark matter candidates. In section~\ref{sec:DM}, we find the relic abundance of each DM particles and identify the constraints coming from DM detection experiments. Some comments about neutrino mass generation are given in section~\ref{sec:neutrino}. Finally, the concluding remarks are presented in Section~\ref{sec:con}. 
\section{ Model}
\label{model}
We study a new physics scenario where the standard model gauge symmetries are augmented by a gauged $\SU$ and a global $U(1)_\X \times \Z_2$.
We supplement the scalar content by two SM singlet scalars: $\phi$ which is a doublet of $SU(2)_\R$:
\beq
\phi = \begin{pmatrix} \frac{1}{\sqrt{2}} (G_\phi^1 + iG_\phi^2)  \\ \\ \varphi^0 +  i G_\phi^3 \end{pmatrix},
\eeq
with $G^i$ being the goldstone bosons, and $\eta$ which is a singlet of $\SU$; both $\phi$ and $\eta$ have non-zero charges under $U(1)_\X$. We also extend the Fermionic fields by a doublet $X_\R =( \chi_2~~~  \chi_1)^T_\R$, and two singlets ($X_{_L} = \chi_1^L, \chi_2^L)$  of $SU(2)_\R$. These fields are complete singlets of the SM gauge symmetries, but they have a non-zero $U(1)_\X \times \Z_2$ charges to avoid mixing with left-handed neutrinos. 

Motivated by the DAMPE excess, we also assume right-handed electron is charged under $SU(2)_\R$. For the notation, we use $E_R =  \left( e' ~~~ e\right)^T_\R $, where $e$ is the familiar SM electron, and $e'$ is a particle with exactly the same quantum numbers as the right-handed electron. The list of the new particles and their charges is presented in Table~\ref{tab:Particle}. 

\begin{table}[h!]
\begin{center}
\begin{tabular}{|c cc cc cc cc|}
\hline 
Particles && $SU(2)_{_R}$ && $U(1)_X$ &&  $\Z_2$ &&$SU(3)_{C}\times SU(2)_{_L} \times U(1)_{Y}  $ \\
\hline
\hline 
$W_{_{R} \mu}$ && 3 && $0$&&   +   &&(1~~~~~,~~~1~~~,~~~~~0)\\
\hline
\hline
$\phi$ && 2 && 1/2 &&   +   && (1~~~~~,~~~1~~~,~~~~~0)  \\
$\eta$ && 1 && 2 &&   --    &&(1~~~~~,~~~1~~~,~~~~~0)  \\
\hline
\hline
$\chi^{1}_{_L}$ && 1 && 1 &&   --   && (1~~~~~,~~~1~~~,~~~~~0)  \\
$\chi^{2}_{_L}$ && 1 && 2 &&  --    && (1~~~~~,~~~1~~~,~~~~~0)  \\
$X_\R $&& 2 && 3/2 &&   --   && (1~~~~~,~~~1~~~,~~~~~0)  \\
$E_{_R}$ && 2 && 1/2&&   +  && $\left(1~~~~~,~~~1~~~,~~-1\right)$\\
\hline
\end{tabular}
\end{center}
\caption{The quantum numbers of the newly introduced particles under the dark symmetries ($\SU \times U(1)_\X \times \Z_2$) and the SM symmetries are presented in this table. }
\label{tab:Particle}
\end{table}%

In the interaction basis, the Lagrangian of the relevant fields has the following form:
\begin{equation}
\cL = \cL_{_{\text{SM}}} + \cL_{_{\text{kin}}} +  \cL_{_{\text{Yuk}}}+ \cL_{_{\text{int}}} - V(H,\phi , \eta),
\end{equation}
where, 
\begin{align}
\cL_{_{\text{kin}} }&=  \frac{1}{4} W^{ a}_{_{R}\mu\nu} W^{ a \mu\nu}_{_R}+\sum_{i = 1,2} \bar{\chi^{i}_{_L}}(\imath \partial\!\!\!/ )\chi^{i}_{_L}  +\bar{X_{_R}}(\imath D\!\!\!/ )X_{_R}+  \bar{E_{_R}}(\imath D\!\!\!/ )E_{_R}\nonumber \\
& +  (D_{\mu}\phi)^{\dagger}(D^{\mu}\phi) + (\partial_{\mu}\eta)^{\dagger}(\partial^{\mu}\eta),\nonumber\\
\cL_{_{\text{Y}}} \ &= \tilde{Y}_{1} \bar{X}_\R  \phi\chi^{1}_{_L} + \tilde{Y}_{2} \bar{X}_\R \tilde{\phi} \chi^{2}_{_L}+\frac{ y_e}{\Lambda}(\bar L  H) ( \phi^\dagger E_\R )  + h.c.,\nonumber\\
 \cL_{_{\text{int}}} &=  \frac{y_{\chi_{L}}}{\Lambda'} (\bar{L^c} \tilde H^\star )(\eta^\star \chi^{^L}_{2})+h.c.
\label{eq:lag} 
\end{align} 
In the kinetic Lagrangian,  $ D_\mu  = \partial_\mu + i g_{_R}  \frac{\tau^a}{2} W_{_{R} \mu}^a + i g_{_Y} B_{_\mu} $, with $g_{_R}$ being the coupling of the $SU(2)_{_R}$, and $g_{_Y}$ is the hypercharge value. The $SU(2)_\R$ field tensor is shown by $W^{\mu\nu}_{_R} = \partial^\mu W_{\R}^\nu - \partial^\nu W_{\R}^\mu -i g_\R [W_\R^\mu, W_\R^\nu]$.  In this Lagrangian,  $\tilde{Y}_i$ are the Yukawa coupling between $\chi_i$ and $\phi$, $\tilde{\phi}= \imath \tau^{2} \phi^\star$ and $\tilde H = \imath \tau^{2} H^\star$. The last term in the Yukawa Lagrangian is the electron Yukawa interaction which due to the charge of $E_\R$ under $SU(2)_\R$ becomes non-renormalizable\footnote{As we will discuss later, $\phi$ acquires a vacuum expectation value and generates a mass for the electrons. The empirical value of electron masses gives a lower bound on $\Lambda$:  $ y_e v_\phi v_h /\Lambda \sim m_e$, which means $\Lambda/y_e  \sim 5 \times 10^ 6 \ \tev$ if  $v_\phi \sim 10 \text{ TeV}$.}.  Another higher dimensional operator that becomes important in figuring out the dynamics of the dark sector is shown in $ \cL_{_{\text{int}}}$. The cut-off scale appearing in $ \cL_{_{\text{int}}}$  does not have to be the same as the one appearing in the electron Yukawa (e.g, $ \Lambda \neq \Lambda'$), and so we distinguish between them. 

To write the scalar potential, $ V(H, \phi, \eta)$,  we first need to comment on whether the new symmetries stay conserved or are broken. To ensure massive gauge bosons and fermions in the dark sector, we assume $\phi$ acquires a vacuum expectation value (vev) and thus breaks the $SU(2)_\R \times U(1)_\X$ at the scale $v_\phi$.  Consequently, the scalar potential becomes\footnote{As it is clear from the form of potential, $\eta$ does not acquire a vev, because its mass terms is positive ($ + \mu_{_\eta}^2$).}
\begin{align}
V(H,\phi ,\eta) &= -  \mu_{_{H}}^2 H^{\dagger}H  - \mu_{_{\phi}}^2 \phi^{\dagger}\phi +\mu_{_{\eta}}^2 \eta^{\dagger}\eta  \nonumber \\ 
&+ \lambda_{_{H}}( H^{\dagger}H)^2  + \lambda_{_{\phi}} (\phi^{\dagger}\phi)^2+\lambda_{_{\eta}} (\eta^{\dagger}\eta )^2 \nonumber \\ 
&+  \xi_{_{H \phi}}( H^{\dagger}H) (\phi^{\dagger}\phi)  + \xi_h ( H^{\dagger}H) (\eta^{\dagger}\eta )+\xi_\phi (\phi^{\dagger}\phi) (\eta^{\dagger}\eta ).
\label{eq:potential}
\end{align} 

Note that since $\phi$ is even under the $\Z_2$ symmetry, the $\Z_2$ symmetry stays conserved after the spontaneous symmetry breaking (SSB). Before moving on to the phenomenological effects of the $SU(2)_\R \times U(1)_\X $ and the Electroweak SSB, we note that the stability of the vacuum puts some constraints on the couplings of the scalar potential~\cite{Kannike:2012pe} 
\begin{align*}
&\lambda_{H} , \lambda_\phi, \lambda_\eta > 0, \hspace{0.2 in} \xi_{H\phi} > -2 \sqrt{\lambda_H \lambda_\phi},  \hspace{0.2 in} \xi_h > -2 \sqrt{\lambda_H \lambda_\eta},  \hspace{0.2 in} \xi_\phi > -2 \sqrt{\lambda_\phi\lambda_\eta},\\
&\sqrt{\lambda_H \lambda_\phi \lambda_\eta} + \xi_{H\phi} \sqrt{\lambda_\eta} + \xi_h \sqrt{\lambda_\phi} + \xi_\phi \sqrt{\lambda_H} \geqslant 0, \\
& \lambda_H \lambda_\phi \lambda_\eta - (\xi_{H\phi}^2 \lambda_\eta + \xi_h^2 \lambda_\phi + \xi_\phi^2 \lambda_H) + 2 \xi_{H\phi} \xi_h\xi_\phi \geqslant 0.
\end{align*}
From minimizing the potential, we can find the values of the vevs:
\beq
v_h ^2 = \frac{4 \lambda_{_\phi}\mu_{_H}^2 - 2 \xi_{_{H \phi}}\mu_{_\phi}^2}{4 \lambda_{_H}\lambda_{_\phi} - \xi_{_{H\phi}}^2},  \hspace{0.2 in} v_\phi ^2 = \frac{4 \lambda_{_H}\mu_{_\phi}^2 - 2 \xi_{_{H \phi}}\mu_{_H}^2}{4 \lambda_{_H}\lambda_{_\phi} - \xi_{_{H\phi}}^2}.
\eeq

One of the most important consequences of the $SU(2)_\R \times U(1)_\X$  and Electroweak SSB is the inducement of the scalar portal. That is the mixing \footnote{Since $\eta$ does not acquire a vev, there is no mixing between the CP-even component of $\eta$ with the other scalars.} between the neutral CP-even component of the Higgs field and that of the $\phi$ field. As a result of this mixing, we have two scalars in the mass basis that interact with both the SM sector and dark sector as a function of the mixing angle $\alpha$. That is 
\beq
\begin{pmatrix} h \\ \varphi \end{pmatrix} = \begin{pmatrix} c_\alpha & -s_\alpha\\ s_\alpha & c_\alpha \end{pmatrix} \begin{pmatrix} h^0 \\ \varphi^0\end{pmatrix},
\eeq
where $h^0$ and $\varphi^0$ are the CP-even component of the Higgs and $\phi$ doublet, respectively, and $h$ and $\varphi$ are the physical fields in the mass basis. We have used $c_\alpha = \cos\alpha$  and $s_
\alpha = \sin \alpha$, with $\alpha$ being 
\beq
\alpha = \frac{1}{2} \tan^{-1} \frac{ \xi_{_{H \phi}} v_h v_\phi}{v_h^2 \lambda_{_H} - v_\phi^2 \lambda_{_\phi}}.\nonumber
\eeq
 The masses of the scalars are, therefore, 
\begin{align*}
m_h^2 &= v_h^2 \lambda_{_H} + v_\phi^2 \lambda_{_\phi} - (v_\phi^2 \lambda_{_\phi} - v_h^2 \lambda_{_H} )/\cos(2 \alpha),\nonumber \\
m_\phi^2 &= v_h^2 \lambda_{_H} + v_\phi^2 \lambda_{_\phi} +  (v_\phi^2 \lambda_{_\phi} - v_h^2 \lambda_{_H} )/\cos(2 \alpha),\nonumber\\
m_\eta^2 &= \mu_\eta^2 + \xi_h v_h^2 + \xi_\phi v_\phi^2 .
\end{align*}

Similarly, we can find the masses of the dark gauge bosons and the fermions:
\beq
m_{_{W_\R}} = \frac{g_\R v_\phi}{\sqrt{2}}, \hspace{0.3 in} m_{\chi_i} = \frac{y_{\chi_i} v_\phi}{\sqrt{2}}  .
\label{eq:masses}
\eeq
One important difference between this symmetry breaking and the EW symmetry breaking is that $U(1)_\X$ is global, and thus does not effect the covariant derivative. Hence, the masses of all of the three gauge bosons associated with $\SU$ ($W_\R$) are the same. 

In this article, we are interested in the phenomenology of the dark matter candidates, and thus it is important to figure out which dark sector particles are cosmologically stable. Given that $SU(2)_\R \times U(1)_\X$ is broken, we need to revisit the conserved symmetries at low scales. Studying the Lagrangian after the SSB, we can convince ourselves that there is a residual $U(1)_\D$ symmetry along with the original $\Z_2$ symmetry, which leads to the stability of at least two particles in the dark sector. The charges of various particles under the $ U(1)_\D \times \Z_2$ symmetry is shown in Table~\ref{tab:candid}, where the $U(1)_\D$ charges are simply $(I_3)_\R + X$, with $X$ being their charges under $U(1)_\X$. 

\begin{figure}
\centering
\includegraphics[width=5cm,height=3.5cm]{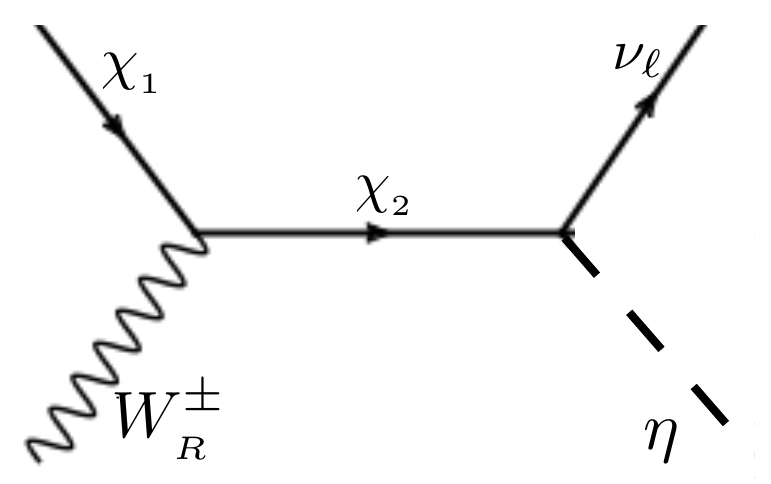}
\caption{Dark matter candidates $(\chi_1, W^\pm_\R, \eta)$ are connected to each other through these diagrams.  }
\label{Fig:decays}
\end{figure}

As $e'$ has electromagnetic charge, it is not a good dark matter candidate. Therefore, we must assume\footnote{A mechanism for $e'$ mass generation is provided in the following subsection~\ref{anomaly}.}  $m_{e'} \gg m_{W_\R}$. Among the other particles listed in Table~\ref{tab:candid}, $W^3_\R$ is also not a DM candidate because it is not charged under either of the $ U(1)_\D\times \Z_2$ symmetries. More specifically, as long as $m_{W_\R} > 2 m_e$ (which as we will show later, the collider constraints require this condition to be true), we can always have the decay of $W^3_\R \to e^+ e^-$.  The rest of the particles mentioned in Table~\ref{tab:candid} are connected through the Feynman diagram shown in Fig.~\ref{Fig:decays}. Depending on the masses of the dark sector particles, they can decay to each other. 
For simplicity, we will assume $m_{\chi_2}$ is considerably larger than the rest of them, so the true players in the DM phenomenology are $\chi_1$, $W^\pm_\R$ and $\eta$. 

\begin{table}[h!]
\begin{center}
\begin{tabular}{ ccccccccc}
\hline 
\hline
 && $\hspace{0.5 in} \Z_2 \hspace{0.5 in}$ &&& $\hspace{0.5 in} U(1)_\D \hspace{0.5 in}$&& \\ 
$\chi_2$ && -- &&& 2 && \\
$\chi_1$ && -- &&&1&& \\
$e'$ &&+&&& 1&&\\
$W_\R^\pm$ && + &&&$\pm1$&& \\
$W_\R^3$ && + &&&0&& \\
$\eta$ && -- &&&2&& \\
\hline 
\hline
\end{tabular}
\end{center}
\caption{The charges of the newly introduced particles under the $ U(1)_D \times \Z_2$ symmetries which are the left over symmetries after the SSB. The lightest particle charged under either of the $U(1)_\D$ and $\Z_2$ symmetries are dark matter candidates. }
\label{tab:candid}
\end{table} 

Collecting the relevant free prameters of our model, we can categorize them into 
\begin{align*}
&\text{Scales}: v_\phi, m_\eta, y_{\chi_L}/\Lambda', \\
& \text{Couplings}: g_\R, y_{\chi_i}, \xi_h, \xi_\phi, \\
&\text{Mixing Angles}: \alpha.
\end{align*}

Particles in the dark sector can interact with SM particles via the scalar portal as well as the direct coupling of the right-handed electrons to dark gauge bosons. In the following section, we first identify the dark matter candidates in each region of the parameter space and then find their relic abundance. We also explain the constraints various experiments impose on the parameter space. However, before diving into the phenomenology, we first address the issue of gauge anomaly that is present in the model.

\subsection{ Anomaly}
\label{anomaly}

The gauged $\SU$ symmetry we have introduced is anomalous. Since gauge anomalies\footnote{ The anomaly in the global $U(1)_\X$, is not dangerous, because the anomalies in global symmetries only lead to the appearance of new vertices.} are dangerous, we need to extend the model to cancel the anomalies.

\begin{itemize}
\item[]1) Among the triangle diagrams, $\SU^3$ is also anomaly free, due to the traceless-ness of the $SU(2)$ symmetries.
\item[]2) The triangle diagram with $\left[\SU\right]^2 U(1)_Y$ would be anomaly-free if and only if the sum of the chiral fermion hypercharges  going through the loop is 0 (e.g., $\sum Y =0$). 
\item[]3) Another triangle diagram that leads to gauge anomaly is  $ \left[Y(1)_Y \right]^3$, which requires the sum of the cube of hypercharges to vanish (e.g., $\sum Y^3 =0$). 
\end{itemize} 

From the points listed above, it is clear that only $E_\R$ leads to gauge anomalies,  because it is charged under both  $SU(2)_\R$ and $U(1)_Y$. The minimal way to cancel the anomalies mentioned in $(2)$ and $(3)$ is to introduce another a doublet of $\SU$ that has hypercharge $+1$, which we call $\Psi= ( \psi_1 ~~~ \psi_2)^T$ , and $\psi_3$ which is a singlet of $\SU$ with $Y= -1$.   

We will have to assume that the mass of $\psi_i$ are large enough that it would not interfere with our phenomenology, but not too large that it would decouple from the theory and leave the model anomalous. To achieve this goal, we will assume there are some vector-like fermions, $f_i$, that can mix with $\psi_i$s after $\phi$ acquires a vev, and thus give $\psi_i$s some mass. Specifically, we will extend our model to include the fermions mentioned in Table~\ref{tab:anomaly}.

\begin{table}[h!]
\begin{center}
\begin{tabular}{|c cc cc cc|}
\hline
Particles && $SU(2)_{_R}$ && $U(1)_Y$&& $U(1)_X$ \\
\hline
\hline 
$E_{_R}$ && 2 &&-1 &&  1/2\\
$\Psi$ && 2&& +1 && $q_\Psi$\\
$\psi_3$ && 1 && -1 && $q_\psi$\\
$f_{e'}$ && 1 && -1 && $1$\\
$f_1$&&1&& 1 && $q_\Psi + 1/2$\\
$f_2$ && 1 && 1 && $q_\Psi - 1/2$\\
$f_3$ && 2 && -1 && $q_\psi+ 1/2$\\
\hline

\end{tabular}
\end{center}
\caption{The quantum numbers of the fermions we need to include in our model to make sure the model is consistent. The $q_k$ with $k= \Psi, \psi$ can be any arbitrary numbers, as long as $q_\psi\neq 0$.  The $f_i$, with $i= 1,2,3,e'$ fermions are vector-like fermions that will mix with the mentioned chiral fermions after $\phi$ gets vev, and lead to a mass for chiral fermions. }
\label{tab:anomaly}
\end{table}%

The Lagrangian terms that lead to a mass for $\psi_i$ and $e'$ are:
\beq
\cL_{\text{VL}} \supset y_{e'} \bar E \tilde \phi f_{e'} +  y_{\psi_1} \bar \Psi  \tilde \phi f_{1} + y_{\psi_2} \bar \Psi \phi f_{2}+ y_{\psi_3} \bar f_3   \phi \psi_{3}  + \sum_{i} M_{f_i} \bar f_i f_i,
\eeq
where $i = e', 1, 2, 3$ in the last Lagrangian term.  We take $M_{f_i}$ to be on the order of $v_\phi$ so that $e', \psi_i$ can acquire a mass at or below $v_\phi$. However, we will assume that these masses are near $m_{\chi_2}$ and thus larger than all of our dark matter candidates. Furthermore, taking $v_\phi \sim O(2 -10\text{TeV})$, we can also be sure that the existence of these particles does not violate the current search on exotic particles with electromagnetism charge~\cite{Schael:2013ita}. It is also noteworthy to mention that we assume there are no vector-like fermions with quantum numbers $(\SU , U(1)_Y, U(1)_\X) = (1, -1, 0)$ to avoid new contribution to the electron mass. 

Having gone over the issue of gauge anomaly, we can now be confident that our theory is consistent. Hence, we can study the phenomenology of DM candidates in the subsequent section. 

\section{Dark Matter Candidates}
\label{sec:DM}

For having a reliable DM model, the DM particles must be long-lived and produce the correct relic density and satisfy the limits of direct and indirect searches. In this section, we examine each of these steps, starting with identify the stable dark sector particles in various regions of the parameter space. 

\begin{figure}
\centering
\includegraphics[width=9cm,height=7.5cm]{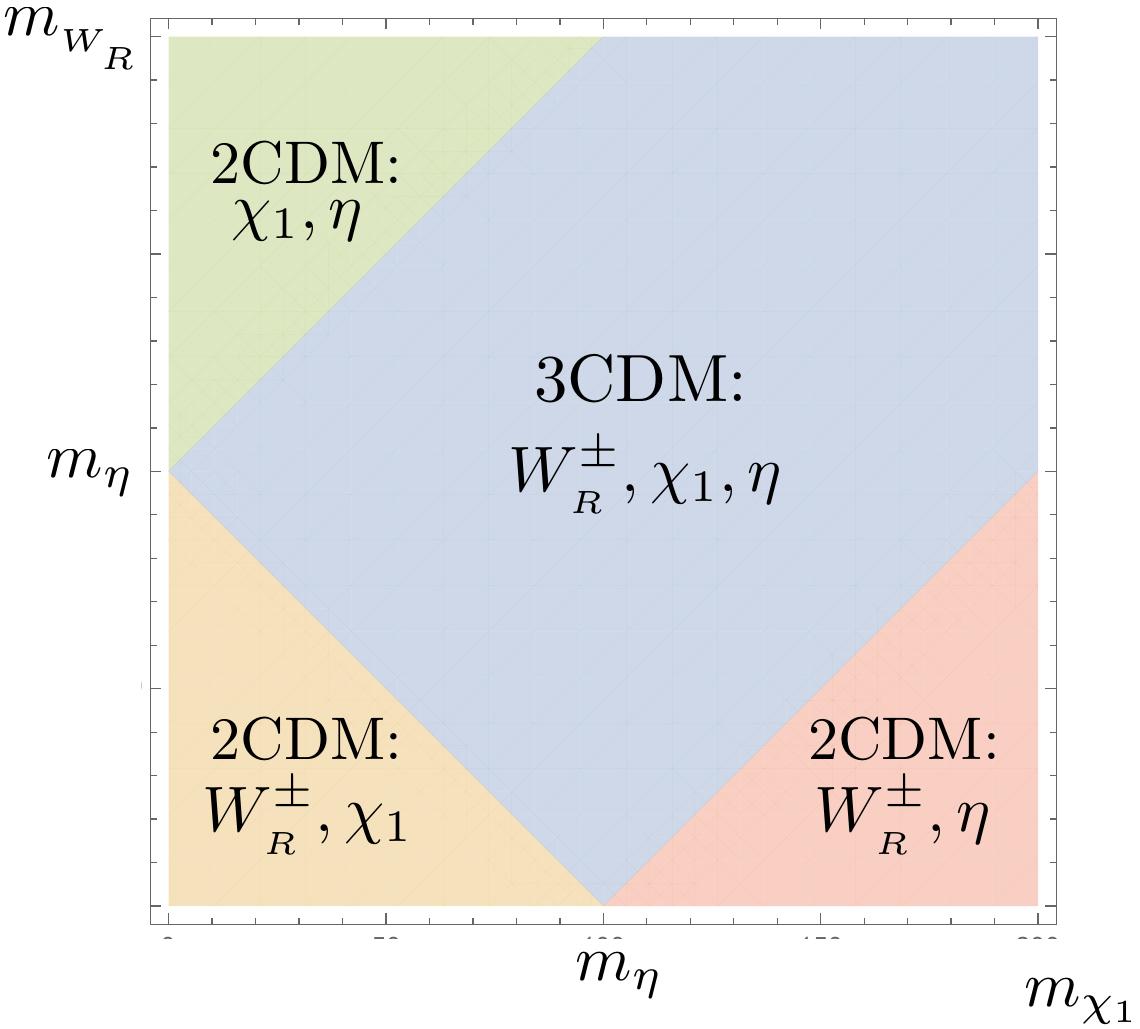}
\caption{The region of parameter space, where the DM candidates are specified. We have assumed $e', ~\chi_2$ are much heavier than $\chi_1, W^\pm_\R,$ and $\eta$ and can decay to lighter dark sector particles.}
\label{Fig:multi}
\end{figure}

\underline{\textit{Stability of the DM candidates}}:

The simplest way to ensure the stability of DM candidates is using the symmetries of the model. There is a $ U(1)_\D\times  \Z_2 $ symmetry that stays conserved after the SSB. Therefore, the lightest particles charged under these symmetries are DM candidate. Taking $e' $ and $m_{\chi_2}$ to be heavier than $\chi_1$, $W_\R$ and $\eta$, we have the following DM candidates:
\begin{itemize}
\item $m_{W_\R} > m_{\chi_{1}}+ m_{\eta}$ : $\eta$ and $\chi_{1}$; 
\item $m_{\eta} > m_{\chi_{1}}+ m_{W_\R}$ :  $W_\R$ and $\chi_{1}$; 
\item $m_{\chi_1} >m_{W_\R}+ m_{\eta}$ : $\eta$  and $W_\R$;
 \item $|m_{\chi_1} - m_{\eta} |<m_{W_\R}< m_{\chi_1}+ m_{\eta}$  : $\eta$, $W_\R$, and $\chi_{1}$,
\end{itemize}
where in the last line we have three DM candidates due to the kinematics. The schematic figures of these conditions are shown in Fig.~\ref{Fig:multi}. In the following subsection, we calculate the relic abundance for each of these DM candidates.

\subsection{Relic Abundance}
\label{sec:RA}

\begin{figure}
\centering
\includegraphics[width=16cm,height=8cm]{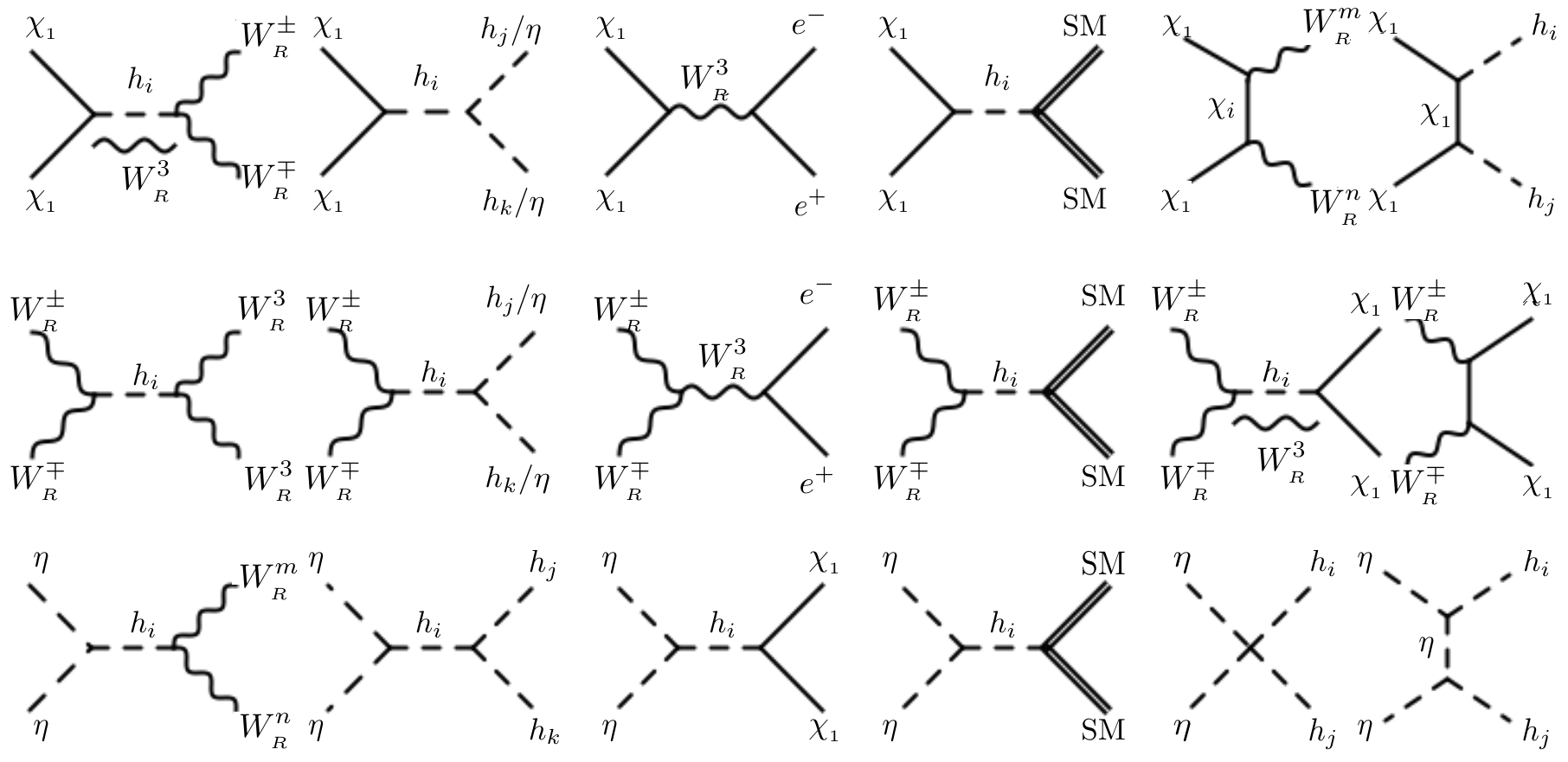}
\caption{The annihilation Feynman diagrams of DM candidates-- Dirac fermion $\chi_1$ (first line), dark gauge boson $W_\R$(second line) and complex scalar $\eta$ (third line) are shown. For notation, we have used $h_i = h, \varphi$, and  SM $ = t, W^\pm,$ and $Z$. For the places where both $\chi_1$ and $\chi_2$ contribute, we have used $\chi_i$, and the places where any of the dark gauge bosons could contribute, we have used $W_\R^{m,n}$. }
\label{Fig:Feynam}
\end{figure}

In thermal Multi-Component Dark Matter (MCDM) scenarios, each dark matter particle starts out in thermal equilibrium with SM particles, and once the temperature falls below the DM mass, DM particles will only annihilate until they freeze-out. The most recent experimental value for relic density $(\Omega_{DM}h^2 =0.119)$ is reported by Planck collaboration~\cite{Ade:2015xua}.
To calculate the DM relic abundance in our model, the coupled Boltzmann equation is applied to study the evolution of the DM particles~\cite{Springel:2005nw}. Assuming thermal relic, we can write:
\beq
\frac{dn}{dt} +3 H n = -\langle \sigma v\rangle (n^2 - n^2_{eq}),
\label{eq:boltz}
\eeq
where $n$ and $n_{eq}$ are denoted the number of density and equilibrium density of the DM particles respectively and $H$ is the Hubbel parameter, and the thermal average annihilation cross section is shown by $\langle \sigma v\rangle$. 
The annihilation Feynman diagrams for all of the DM components are depicted in Fig~\ref{Fig:Feynam}, where SM denotes $W, Z$ bosons and the top quark. 

For Eq.~\ref{eq:boltz} to be valid, we need to make sure $\Gamma_{\chi_2} \gg m_{\text{DM}}$. In other words, we want $\chi_2$ to decay long before the DM particle become non-relativistic. Therefore, 
\begin{equation*}
\Gamma_{\chi_2} \sim \frac{ y_{\chi_L}^2 m_{\chi_2}^3}{\Lambda'^2 16 \pi ^2}  \gg H = \left. \frac{ 1.66 \sqrt{g_*^\rho}\  T^2},{M_{\text{Pl}}}\right|_{T = m_\text{DM}},
\end{equation*}
where $M_{\text{Pl}}$ is the reduced Planck mass and $g_\star$ represents the relativistic degrees of freedom at temperature $T$. This constraint puts a mild bound on $\left(\Lambda'/y_{\chi_ L}\right)  \lesssim \sqrt{ M_{\text{Pl}} \ m_{\chi_2}^3}/ (12 \pi m_{\text{DM}}) $. For example, if we care about DM particles with $O(\text{TeV})$ mass, and so we take $m_{\chi_2}\sim 10\  \text{TeV}$ and $y_{\chi_L} \sim O(1)$, we get  $ \Lambda' < 5 \times10^{7} \ \text{TeV}$.  Furthermore, we need to assume any of the $W^\pm_\R,\, \chi_1,$ or $\eta$ that is not DM decay quickly enough that they do not interfere with the relic abundance of DM particles once DM becomes non-relativistic. Hence, if we show the decaying particle by $DM'$, we roughly get 
\begin{equation*}
O(1) \frac{ m_{DM'}^5}{\Lambda'^2 m_{\chi_2}^2 16 \pi ^2}  \gg H = \left. \frac{ 1.66 \sqrt{g_*^\rho}\  T^2}{M_{\text{Pl}}}\right|_{T = m_\text{DM}},
\end{equation*}
where $O(1)$ represent the couplings $ y_{\chi_L}$ and the other couplings involved. Taking $m_{DM'} \sim 2\ \text{TeV}$, and using the same benchmarks as before, we arrive at a slightly more stringent bound on $\Lambda' < 10^{6} \ \text{TeV}$. As long as this condition is satisfied, we can be confident that the decays of heavier dark sector particles do not play a role in the relic abundance of DM candidates. 

The only diagram that leads to semi-annihilation between DM candidates is the one shown in Fig.~\ref{Fig:decays}, which is roughly 
\begin{equation*}
\langle \sigma v \rangle_{DM_1 DM_2 \to DM_3 \nu_\ell}\sim \frac{ g_\R^2  y_{\chi_L}^2 m_{DM}^2}{ 32 \pi \Lambda^{'2} m_{\chi_2}^2}, 
\end{equation*}
where $DM_i = \chi_1, W^\pm_\R, $ and $\eta$, and we have assumed all of them have roughly the same mass, $m_{DM}$. Using the usual benchmark values: $g_\R \sim  y_{\chi_L} \sim 1, \  m_{DM} \sim 1 \ \text{TeV}, m_{\chi_2} \sim 10 \ \text{TeV},$ and taking\footnote{In section~\ref{sec:neutrino}, where we discuss neutrino mass generation, we find that $\Lambda'$ should preferably be bigger than $10^4 \ \text{TeV}$.}  $\Lambda' \sim 10^4 \ \text{TeV}$, the semi-annihilation cross section is approximately $10^{-37} \text{cm}^2$, and thus is extremely small. Therefore, we ignore the semi-annihilation diagrams. Consequently, the calculation of relic abundance is greatly simplified and the only important ingredient we need is the annihilation cross sections of each of the DM candidates. The analytical expressions of the annihilation diagrams can be found in Appendix~\ref{app:xsec}~\cite{Berlin:2014tja, Ko:2014gha}.

\begin{figure}
\centering
\includegraphics[width=17cm,height=6cm]{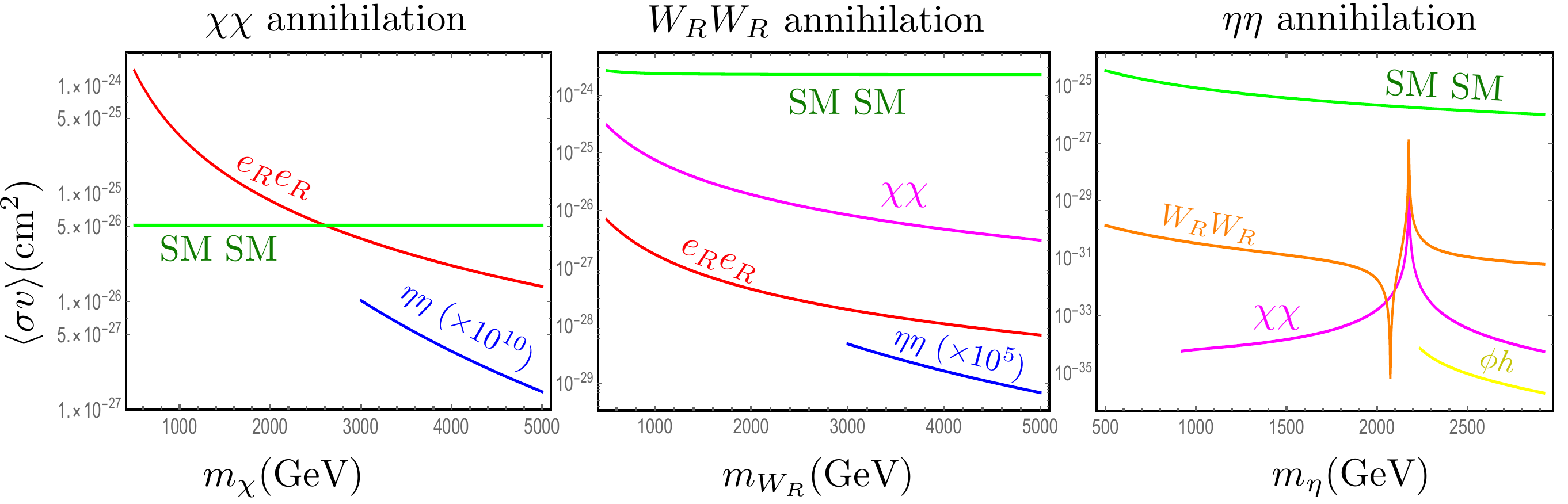}
\caption{The annihilation of DM candidates to various final states, where we have fixed $g_\R = 0.65, \ \yc = \xi _h = \xi_\phi = 0.3$, $ s_\alpha = 0.1$, $ m_\phi = v_\phi$, and $ m_{\chi_2} = 10 \ \text{TeV}$. The {\bf left panel} is the annihilation of $ \chi \chi$, taking $m_\eta = 3 \ \text{TeV}$; the {\bf middle panel} shows the cross section rates of $W_\R W_\R$ annihilations, where we have again taken $m_\eta =3 \ \text{TeV}$; and finally the {\bf right panel} illustrates the annihilation of $ \eta \eta$, where $m_{W_\R}$ is fixed to $ 2 \ \text{TeV}$. The red line is the annihilation to a pair of right-handed electrons through s-channel $W^3_\R$ mediation. The green line indicates the annihilation to SM particles $= t, W^\pm, Z, h$. The magenta line is the annihilation of DM candidates to $\chi \chi$, the orange line is the annihilation to a pair of $W_\R$ and the blue line is when the DM particle is annihilating to $\eta \eta$. Because the annihilations to $\eta \eta$ are very small, their cross sections were multiplied by $ 10^{10}$ (left panel) and $10^5$ (middle panel).  }
\label{Fig:annihilatins}
\end{figure}

To get a better understanding of the relative sizes of these annihilation process with respect to each other, Fig.~\ref{Fig:annihilatins} shows the cross section of various diagrams where we have fixed: $g_\R = 0.65, \ \yc = \xi _h = \xi_\phi = 0.3$, and $ s_\alpha = 0.1$. We have also fixed $ m_\phi = v_\phi$ and $ m_{\chi_2} = 10 \ \text{TeV}$. The left panel of Fig.~ \ref{Fig:annihilatins}, shows the annihilation $\chi \chi$ to various final states. The red line is $ \chi \chi \to e_\R e_\R$ though $ W^3_\R$, and as we can see it has a very significant rate. $ \chi \chi \to  \text{SM SM}$, where $SM = t,W^\pm, Z, h$ is shown in green. The blue line is the cross section of $ \chi \chi \to \eta \eta$ times a factor of $10^{10}$, where we have taken $m_\eta = 3 \ \text{TeV}$. This channel opens up for $m_\chi > 3 \ \text{TeV}$ and the rate is very small $(\sim 5 \times 10^{-37} \text{cm}^2)$.  With the parameters chosen, $m_{\chi}$ is smaller than $m_{W_\R}$ and $m_\phi$ (Eq.~\ref{eq:masses}), and thus the annihilation of $ \chi \chi \to W_\R W_\R, \phi h, \phi \phi$ does not happen. The middle panel of Fig.~ \ref{Fig:annihilatins} shows the annihilation of $W_\R W_\R$ to various final states, where we have again taken $m_\eta = 3 \ \text{TeV}$. The annihilation of $W_\R W_\R \to e_\R e_\R$ is p-wave and thus it is comparatively smaller than $ \chi \chi \to e_\R e_\R$. The annihilation of $ W_\R$s to SM particles and $\eta$, however, benefit from a higher coupling ($ g_\R > \yc$) and thus it is relatively bigger. Furthermore, the annihilation of $W_\R$ to a pair of $\chi$s is also kinematically possible and has a fairly large rate\footnote{In the region where both $W_\R$ and $\chi_1$ are DM candidates, the Boltzmann equation becomes coupled due to the annihilation of $W_\R W_\R \to \chi_1 \chi_1$, and needs a more careful treatment. However, due to the much smaller rate of this channel compared with $W_\R W_\R \to \text{SM SM}$, and the mass difference between $W_\R$ and $\chi_1$, we noticed that annihilation of $W_\R$ to $\chi_1$ does not play a significant role.}. Finally, the right panel of Fig.~ \ref{Fig:annihilatins} illustrates $\eta$ annihilation, where we have fixed $m_{W_\R} = 2  \ \text{TeV}$. The resonance at around $2 \ \text{TeV}$ is due to $\phi$ becoming on-shell in s-channel annihilations of $ \eta$. The yellow line is the annihilation of $\eta \eta \to \phi h$ which opens us for $ 2 m_{\eta} > m_\phi+ m_h$. Other than $\eta \eta \to \text{SM SM }$, the rest of the channels suffer from low rate.

Having determined the important processes that set the relic abundance of DM, we move to current constraints on the model parameters. In the following two sub-sections, we study the direct detection, indirect detections as well as the collider constraints. We show that if we insist on using $O(1)$ couplings, the allowed parameter space can be probed with the next generation of experiments.

\subsection{Direct Detection}
\label{sec:DD}

Since, in MCDM, each DM particle shares some portion of the total relic abundance, we expect their annihilation rate to be larger than what would be single component DM $\sim 2.2 \times 10^{-26} \text{cm}^3/\text{s}$:
\begin{align*}
 \Omega_{\DM_1} + &  \Omega_{\DM_2} +\cdots +  \Omega_{\DM_n} = \Omega_{\DM_{\text{total}}} \\
 &  \Rightarrow  \Omega_{\DM_1}  <\Omega_{\DM_{\text{total}}} \Rightarrow   \langle\sigma v\rangle_{\DM_1} >  2.2 \times 10^{-26} \text{cm}^3/\text{s}
 \end{align*}
 Naively, there is a concern that with such large interaction rates of DM with SM particles, it must have been detected at DM experiments, by now. One of the most important constraints on DM models comes from Direct Detection (DD). In our model, DM can scatter with nucleus through Higgs or $\phi$ exchange, leading to potential constraints from DD. Since Higgs portal interactions care about the mass of particles, the interaction of DM with the nucleon is suppressed. In other words, Higgs portal scenarios are efficient in producing the right relic abundance through the annihilation of DM to heavy SM particles, but have a suppressed scattering cross section in DD experiments. This particular reason is common to all Higgs portal DM, and it is one of the benefits of the Higgs portal over generic $Z'$ models. 
 
Furthermore, in calculating the relic abundance of $ \chi_1$ and $W_\R^\pm$ DM, their annihilation to a pair of electrons through $W_\R^3$ mediator is dominant, especially for large values of $g_\R$. However, this process contributes to DD only at loop level and thus is negligible. This is the second reason that we can have efficient annihilation of $\chi_1$ and $W_\R^\pm$ DM while being safe from DD bounds. 

Since the mediator is a CP-even scalar, the bounds on our model comes from spin-independent DD. Higgs portal DD constraints have been studied in multiple studies, and it is well-known that if DM is a Dirac fermion, $\chi$, then its scattering cross section with the SM is~\cite{Barger:2007im}
\beq
\sigma_{{\chi_1{\text{-N}}}} = \frac{ y_\chi^2 \sin^2 2 \alpha}{4\pi} m_{\text{red}}^2 \left(\frac{1}{m_h^2} - \frac{1}{m_\phi^2}\right)^2 g_{Hp}^2,
\label{eq:diracdd}
\eeq
where $g_{_\DM}$ is the coupling of the DM particle with the scalar mediator,  $m_{\text{red}} = m_p m_\DM / (m_p + m_\DM),$ and  $g_{Hp}$ is the effective coupling of Higgs with proton~\cite{ Djouadi:2012zc}:
\beq
g_{Hp} = \frac{m_p}{v_h} \left[ \sum_{q = u,d,s} f_q^{(p)} + \frac{2}{9} \left(1- \sum_{q = u,d,s} f_q^{(p)}\right)\right] \approx 1.3 \times 10^{-3}.
\eeq

In the case the dark matter is a gauge boson, its scattering cross section with nucleons goes as the following\footnote{As shown in Eq.~\ref{eq:diracdd}, and Eq.~\ref{eq:gaugedd}, there is a destructive interference between the two scalar mediation in DD bounds for the case of Dirac fermion and gauge boson DM which is another reason that DD cannot bound Higgs portal DM models very well.}~\cite{Baek:2012se} :
\beq
\sigma_{_{W_\R\text{-N}}}= \frac{ g_R^2 \sin^2 2 \alpha}{4 \pi } m_{\text{red}}^2 \left(\frac{1}{m_h^2} - \frac{1}{m_\phi^2}\right)^2 g_{Hp}^2,
\label{eq:gaugedd}
\eeq
and finally for a stable scalar it is~\cite{Casas:2017jjg}
\beq
\sigma_{_{\eta{\text{-N}}}} = \frac{m_p^2}{8 \pi m_\eta^2 } m_{\text{red}}^2 \left(\frac{\xi_h c_\alpha}{m_h^2} + \frac{\xi_\phi s_\alpha }{m_\phi^2}\right)^2 g_{Hp}^2.
\eeq

To recast the DD bounds on our model, it is important to realize that each component of DM constitute only a percentage of DM. Assuming their ratio in early universe is the same as the one in the vicinity of earth, we get 
\beq 
\sigma_{_{\DM \text{-N}}}= \sigma_{{\chi_1{\text{-N}}}} \times \frac{ \Omega_{\chi_1}}{\Omega_{\DM}} + \sigma_{_{W_\R\text{-N}}} \times \frac{ \Omega_{W_\R}}{\Omega_{\DM}} + \sigma_{_{\eta{\text{-N}}}} \times \frac{ \Omega_\eta}{\Omega_{\DM}}.
\eeq

\begin{figure}
\includegraphics[width=17 cm,height=8cm]{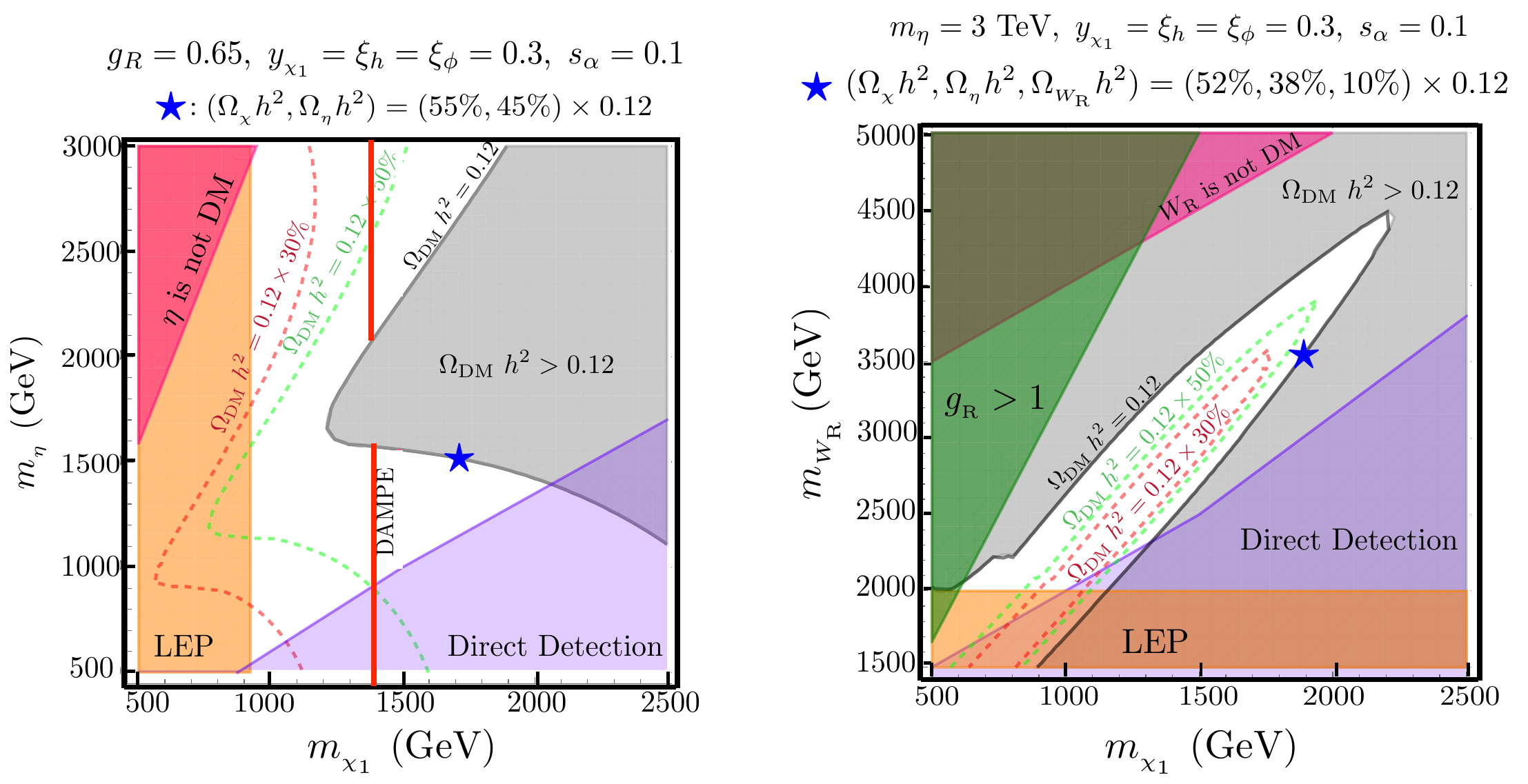}
\caption{The allowed parameter space for various benchmarks. The solid black dashed green and dashed red shown the contour for producing all, $50\%$, and $30\%$ of the relic abundance of DM, respectively. The shaded gray region is when we over-produce DM. The shaded purple is excluded from DD experiment~\cite{Messina:2018fmz}, and the shaded orange part is excluded by LEP~\cite{Schael:2013ita}.  The pink region is when the indicated particle is no longer DM and decays. The green region is when $g_\R >1$ which violates perturbativity. The star is chosen as an example to show how much of DM each particle constitutes. Finally, the solid red line is the region we can explain the DAMPE excess~\cite{Ambrosi:2017wek}. }
\label{Fig:bench}
\end{figure}

In Fig.~\ref{Fig:bench}, the DD constraints as well as some other constraints are shown. The purple region is excluded from the DD experiments~\cite{Messina:2018fmz}. The LEP experiment puts a stringent constraint on any particle that interacts with electrons~\cite{Schael:2013ita}. Since right-handed electrons are charged under $SU(2)_\R$, the dark gauge bosons can directly interact with them. The strongest constraint of LEP comes from the Drell-Yan Production of a pair of electrons through an exchange of $W_\R^3$, which excludes $m_{W_\R} < 2 \ \text{TeV}$. This is shown in orange in Fig.~\ref{Fig:bench}. The red shaded region is when the indicated particle is no longer a DM candidate because it is not stable. The green region is when $g_\R >1 $, which threatens perturbativity. Finally, the gray region is when the relic abundance of all DM candidates combined is too large and they over-close the universe. The green and red dotted lines indicate that the DM introduced in this paper are respectively $50\% $ and $30\%$ of the total DM. The star in the left plot of Fig.~\ref{Fig:bench} is a benchmark, where $55\%$ of the DM is due to the relic abundance of $\chi$ and $45 \%$ is from $\eta$. Similarly, the star in the right plot of  Fig.~\ref{Fig:bench}, indicates a sample point, where $\chi$, $\eta$ and $W_\R$ are respectively $52\%$, $38 \%$ and $10 \%$ of the total DM. Due to the large cross-section of $W_\R$ to electrons and $\phi$, its relic abundance is usually low. 

\subsection{Indirect Detection}
\label{sec:ID}

Another way to constrain our parameter space is by using indirect detection (ID) results. The main annihilation channels of our DM candidates are the production of a pair of electrons or heavy particles. Heavy particles eventually decay to stable particles, which some of them can be detected here on earth. Furthermore, any particle in this process that is electromagnetically charged will radiate photon which can also be detected through various experiments (e.g, Fermi-LAT~\cite{Ackermann:2013yva}). However, due to the large uncertainty of the background, ID bounds are usually mild.  Even considering the strongest bounds of Fermi-LAT, which is $100\%$ branching ratio to $b \bar b$, ID can constrain DM only up to a few hundreds of GeV, which is smaller than the benchmarks we are considering. 

Recently, the DArk Matter Particle Explorer (DAMPE) experiment~\cite{Ambrosi:2017wek}, which is a satellite-borne, high energy particles and gamma-ray detector, published their measurement of the electron plus positron spectrum. Their result indicates a tentative narrow peak around $\sim 1.4 \ \tev$. The local significance of this excess is about $3.7 \sigma$ assuming a broken power-law background \cite{Fowlie:2017fya,Huang:2017egk}, and its global significance has been measured to about $2.3 \sigma$ \cite{Huang:2017egk,Niu:2017lts,Ge:2017tkd,Chao:2017emq,Nomura:2018jkd,Yao:2018ewe}. Such a narrow peak could be a result of a DM with mass $1.4 \ \tev$ to a pair of right-handed electrons. The interaction of DM with left-handed electrons should be suppressed, due to the results published by IceCube, which reported no excess in the neutrino experiment~\cite{Zhao:2017nrt}. This is the reason we considered only the right-handed electrons being charged under the $SU(2)_\R$. 
 
 According to the DAMPE experiment, the annihilation rate to electron-positron is estimated to be much more than the annihilation rate for a single component DM, which further motivates our set up for multi-component DM. However, it is important to make sure the annihilation to electron pair is s-wave. 
 
Among the DM candidates in our set-up,  $W_\R^\pm$ and $\chi_1$ interact with right-handed electrons strongly. The annihilation of $W_\R^\pm$ to $e^+ e^-$, however, is p-wave:
\beq
 \langle \sigma v \rangle _{_{W_\R W_\R \to e^+ e^-}} = \frac{ 5\  g_\R^4}{216 \pi m_{W_\R}^2} v^2. 
\eeq
Even though this process could play a significant role in setting the abundance of $W_\R^\pm$ in the early universe, its rate right now should be negligible. That is because the ambient velocity of DM is estimated to be $ v_\DM \sim 0.001$. The annihilation of $\chi_1 \chi_1 \to e^+ e^-$, on the other hand, is s-wave and thus can have a significant contribution to ID at the current time:
\beq 
 \langle \sigma v \rangle_{_{\chi_1 \chi_1 \to e^+ e^-}} = \frac{ 8 \ g_\R^4 m_{\chi_1}^2 }{64 \pi (m_{W_\R}^2 - 4 m_{\chi_1}^2)^2}.
\eeq 
Therefore, in the region where $\chi_1$ is a DM  particle, its annihilation to a pair of electrons could justify the observation of the narrow peak in the DAMPE experiment. The red line in the left panel of Fig.~\ref{Fig:bench}, shows the benchmark that could explain the DAMPE observation.  It is worth mentioning that even though the main motivation behind charging the right-handed electrons under $SU(2)_\R$ was explaining the DAMPE observation, the annihilation of DM particles to a pair of electrons contributes significantly in setting the relic abundance of DM candidates. In the scenario where right-handed electrons did not interact directly with the dark sector, DM particles had to be about a factor of 5 lighter to not over-close the universe. However, that region of the parameter space is strongly constrained by DD experiments.

\subsection{Neutrino Mass}
\label{sec:neutrino}
\begin{figure}
\includegraphics[width=5 cm,height=4 cm]{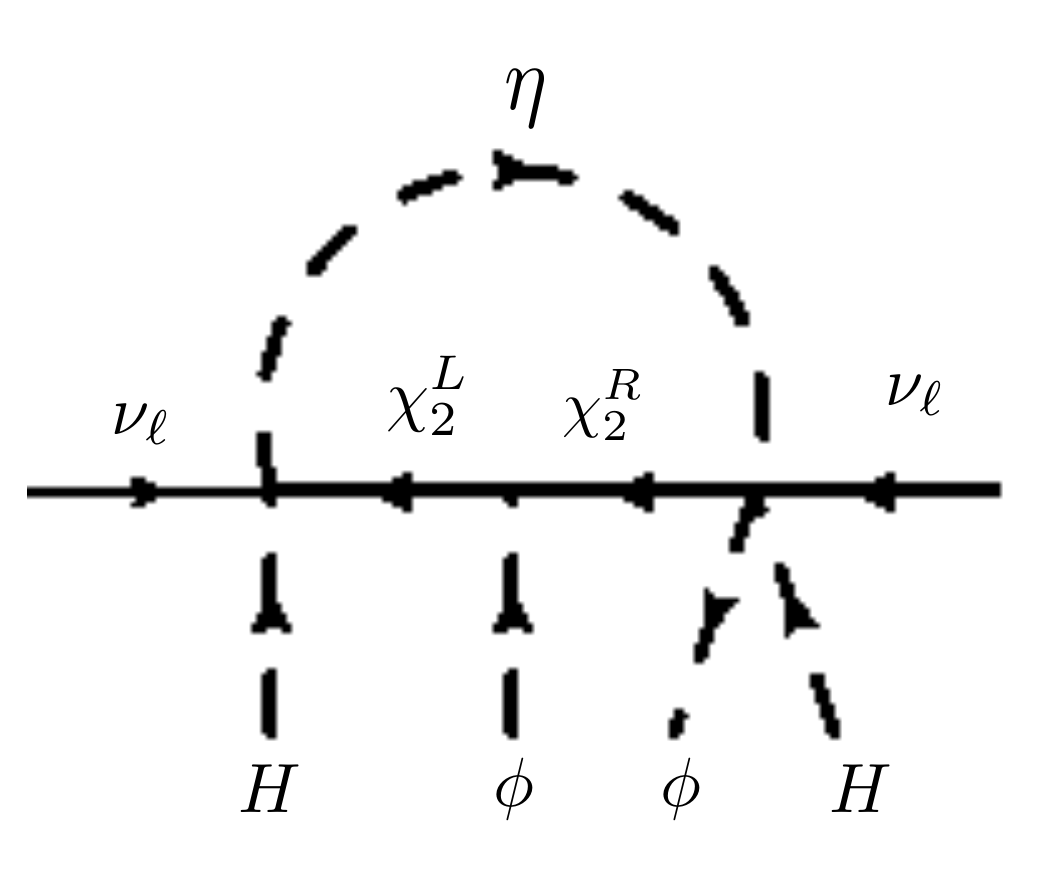}
\caption{The radiative generation of mass of neutrino through non-renormalizable operators.}
\label{Fig:neutrino}
\end{figure}

An added bonus of non-minimal structure in the dark sector is that we can attack some other problem of the SM. In this part, we comment on how the neutrino mass can be radiatively generated using the particles in the dar sector. To do so, we will employ the following terms:
\beq
\frac{y_{\chi_L}}{\Lambda'} (\bar L^c \tilde H^\star)(\eta^\star \chi_2^L) + \frac{y_{\chi_\R}}{\Lambda^{'2}} (\eta \bar X_\R \tilde \phi) (\tilde H^\dagger L) + \tilde Y_2  \bar X_\R \tilde {\phi} \chi_2^L
\eeq
We can think of $\Lambda'$s to be vector-like fermions, one with charge $(SU(2)_L,  U(1)_Y, U(1)_X, \Z_2)$ $= ( 2, -1/2, 2, -)$ and another with charge $(SU(2)_L, U(1)_Y, U(1)_X, \Z_2) = ( 1, 0, 2, -)$.  To avoid the contribution of Weinberg operator in giving neutrinos a mass, we will assume there are no Majorana or triplet of $SU(2)_W$ in the UV theory\footnote{Even though the Weinberg operator respects the symmetries of the model, since it violates lepton number, it requires a new degree of freedom in the UV to generate the term. In other words, we cannot generate the Weinberg operator with the degrees of freedom present at low scales. Thereby, we can ignore the effect of Weinberg operator by requiring there to be no degrees of freedom in the UV that can generate such term.  It is noteworthy to mention that we cannot impose the lepton number, $U(1)_L$, as a symmetry of the model, because one of the higher dimensional operators we used to generate the neutrino mass term violates $U(1)_L$.}. The diagram leading to neutrino mass is shown in  Fig.~\ref{Fig:neutrino}.

Given that neutrino mass is expected to be smaller than a few 0.1 eV \cite{Giusarma:2016phn,Giusarma:2018jei} we can roughly estimate the value of $\Lambda'$ assuming $y_{\chi_L}$ and $y_{\chi_R}$ are $O(1)$:
\beq
\frac{v_h^2 v_\phi^2  m_{\chi_2}^2}{\Lambda^{'3} m_\eta^2 16 \pi^2}  \lesssim 0.1\  \text{eV}.
\eeq 
 Assuming a benchmark value of $v_\phi \sim 10 \ \text{TeV}$,  $ m_{\chi_2} \sim 10 \ \text{TeV}$, and $\eta \sim 3\ \text{TeV}$  we get $ \Lambda' > 2 \times 10^ 4 \ \text{TeV}$.  This constraint combined with the bound we need to satisfy to make sure the decaying particles decay before DM candidates become non-relativisitc requires $\Lambda'$ to be roughly in the range of $10^4 -10^6$ TeV.
\section{Conclusions}
\label{sec:con}

In this work, we studied a scenario of the dark sector that contains two or three dark matter (DM) candidates. We proposed extending the SM symmetries by a gauge $SU(2)_\R$ and a global $U(1)_\X\times \Z_2$, which the continuous symmetries are spontaneously broken to a global $U(1)_D$. We considered a case where dark sector contained some Dirac fermions and complex scalars to investigate a dynamic dark sector. To see how our proposed scenario could explain the recent observation by DArk Matter Particle Explorer (DAMPE), we also charged right-handed electrons under $SU(2)_\R$. We assumed $O(1)$ couplings, to consider a more natural scenario. Other than the Higgs portal, which connects the dark sector to the SM, the annihilation of $W_\R$ and $\chi_1$ to a pair of electrons happen to play a significant role in the relic abundance of DM particles. 

The phenomenology of DM candidates was studied, and the region of the parameter space where they can produce the right relic abundance while being safe from various DM detection experiment was identified. We noticed that only a small region of parameter space survives the constraint and this region could be probed with the next generations of experiments. Additionally, we commented on how neutrinos can gain a mass through non-renormalizable interactions with the dark sector. An important advantage of our scenario over Weinberg operator is that our cut-off scale is $O(10^4 \ \text{TeV})$, and much lower than the cut-off scale suggested by the Weinberg operator.  

In conclusion, we emphasize that in the era where single DM thermal relics are highly constrained, it is important to consider multi-species DM. In the most simplistic paradigm, where DM particles are thermal relics, multi-component DM suggests strong couplings between DM particles and SM. As a result, Leptophilic DM or when there is a Higgs portal models are preferred.

\section*{Acknowledgments}
\label{sec:ack}

We would like to especially thank Adam Martin for his invaluable comments on the draft. We are also grateful to Carlos Alvarado, Hoda Hesari, Mojtaba Mohammadi, and Sedigheh Tizchang for insightful discussions. We  thank the CERN theory group for their hospitality. 

\appendix 
\section{The cross section of DM candidates}
\label{app:xsec}
In this appendix, we show the analytical expressions that we have calculated using FeynCalc~\cite{Shtabovenko:2016sxi}. The first subsection is the potential annihilation cross sections of $\chi$, the second one belongs to $ W_\R$ and the last one shows the annihilation cross sections of $\eta$. These processes set the relic abundance of DM if they are 1) kinematically allowed, 2) the indicated initial state is indeed a DM candidate. 
\subsection{$\chi$ DM }
\label{sec:chi}
\begin{scriptsize}
\begin{align*}
&\langle \sigma v \rangle_{\chi \chi \to e_\R e_\R} = \frac{ 8 g_\R^4 m_\chi^2}{64 \pi (4m_\chi^2 - m_{W_\R}^2 )^2}\\
&\langle \sigma v \rangle_{\chi \chi \to t \bar t} = \frac{9  y_t^2 \yc^2 }{8 \pi} c_\alpha^2 s_\alpha^2  \frac{(m_h^2 m_\phi^2)^2 (m_\chi^2 - m_t^2)\sqrt{4 m_\chi^2 - m_t^2}}{m_\chi (m_h^2 - 4m _\chi^2)^2 (m_\phi^2 - 4m_\chi^2)^2} v^2\\
&\langle \sigma v \rangle_{\chi \chi \to W W} = \frac{\yc^2  c_\alpha^2 s_\alpha^2}{8 \pi} \frac{ (m_h^2 - m_\phi^2)^2 (3 m_w^4 + 4m_\chi^4 -4 m_\chi^2 m_w^2) \sqrt{ 4 m_\chi^2 - m_w^2}}{v_h^2 m_\chi (m_h^2 - 4m_\chi^2)^2 (m_\phi^2 -4 m_\chi^2)^2} \\
&\langle \sigma v \rangle_{\chi \chi \to ZZ} = \frac{\yc^2  c_\alpha^2 s_\alpha^2}{2 \pi} \frac{ (m_h^2 - m_\phi^2)^2 (3 m_z^4 + 4m_\chi^4 -4 m_\chi^2 m_z^2) \sqrt{ 4 m_\chi^2 - m_z^2}}{v_h^2 m_\chi (m_h^2 - 4m_\chi^2)^2 (m_\phi^2 -4 m_\chi^2)^2}\\
&\langle \sigma v \rangle_{\chi \chi \to hh } =  \frac{\yc^2 s_\alpha^2}{64 \pi  m_\chi^3  v_h^2 (4m_\chi^2 - m_h^2)^{3/2} (4 m_\chi^2 - m_\phi^2)}\\
& \hspace{1 in}\left[ (18 c_\alpha^2 m_h^4 m_\chi^2 (m_h^2 - m_\phi^2)^2\right. \\
& \hspace{1 in}+  24 \yc v_h s_\alpha c_\alpha m_h^2 sm_h^2m_\chi (4 m_\chi^2 - m_h^2 ) (m_\phi^2 - m_h^2)(4m_\chi^2 - m_\phi^2)\\
& \hspace{1 in}+ \left. 4 \yc^2 s_\alpha^2 v_h^2 (4m_\chi^2 - m_h^2) (4m_\chi^2 - m_\phi^2)\right]\\
&\langle \sigma v \rangle_{\chi \chi \to \phi \phi } =  \frac{\yc^2 c_\alpha^2}{32 \pi  m_\chi  v_\phi^2 (4m_\chi^2 - m_h^2)^{2} (2 m_\chi^2 - m_\phi^2)(4 m_\chi^2 - m_\phi^2)^{3/2}}\\
& \hspace{1 in} \left[  128\,  \yc^2 m_\chi^6 v_\phi^2 -12 m_\chi m_\phi^6 (3m_\chi + v_\phi \yc)  \right.\\
& \hspace{1 in}+ 4 m_\chi^2 m_\phi^4(9 m_\chi^2+ 18 m_\chi v_\phi \yc + 2 v_\phi^2 \yc^2)\\
& \hspace{1 in}-32 \yc m_\chi^4 m_\phi^2 v_\phi (3m_\chi + 2 v_\phi \yc)\\
& \hspace{1 in}- 6 c_\alpha^2 s_\alpha^2 m_\phi^2(4m_\chi^2 - m_h^2)(8 m_\chi^4 + m_\phi^4- 6 m_\chi^2 m_\phi^2)(8 m_\chi^3 v_\phi\yc + 3m_\phi^4 - 2m_\chi m_\phi^2(3m_\chi+ v_\phi\yc))\\
& \hspace{1 in}+\left. 9 s_\alpha^4 m_\phi^4(8 m_\chi^4 + m_\phi^4 - 6m_\chi^2 m_\phi^2)^2\right] \\
&\langle \sigma v \rangle_{\chi \chi \to h \phi } =  \frac{s_\alpha^2 \yc^2 \sqrt{ m_h^4 + (4 m_\chi^2 - m_\phi^2)^2 - 2 m_h^2 (m_\phi^2 - 2 m_\chi^2)}}{512 \pi  v_h^2 m_\chi^4 (4 m_\chi^2 - m_h^2)^2} (4m_\chi^2 - m_\phi^2)^2 (4 m_\chi^2 - m_\phi^2 - m_h^2)^2\\
 & \hspace{1 in} \left[18 s_\alpha^2 m_h^4 m_\chi^2 (4 m_\chi^2 - m_\phi^2)^2 ( 4 m_\chi^2 - m_\phi^2 - m_h^2)^2 +(m_h^2 (25 m_\phi^4 - 128 m_\phi^2 m_\chi^2 + 256 m_\chi^4) \right.\\
 & \hspace{1 in} - 6 s_\alpha c_\alpha \yc  v_h^2 m_h^2 m_\chi (4 m_\chi^2 - m_h^2) (4 m_\chi^2 - m_\phi^2) (4 m_\chi^2 - m_\phi^2 - m_h^2)(3 \sqrt{2} m_h^2 m_\phi^2 - (4 m_\chi^2 - m_\phi^2) ( 3 \sqrt{2} m_\phi^2 + 16 m_\chi^2))\\
  & \hspace{1 in}+ c_\alpha^2 (4m_\chi^2 - m_h^2)^2 (2 m_h^2 m_\phi^2 (4 m_\chi^2 - m_\phi^2) ( 7 m_\phi^2 - 8 ( 8 + 3 \sqrt{2}) m_\chi^2)\\
    & \hspace{1 in} \left. + \yc^2 v_h^2 ( 4m_\chi^2 - m_\phi^2)^2 (25 m_\phi^4 + 16 (3 \sqrt{2} - 8) m_\phi^2 m_\chi^2 + 384 m_\chi^4)\right]
 \end{align*}
\end{scriptsize}

\begin{scriptsize}
\begin{align*}
 &\langle \sigma v \rangle_{\chi \chi \to W_\R W_\R} = \frac{ \sqrt{ 4m_\chi^2 - m_{W_\R}^2}}{256 \pi m_\chi} \left\{ \frac{160 \yc^2 ( 4 m_\chi^2 + 3 m_{W_\R}^4  4 m_\chi^2 m_{W_\R}^2) ( c_\alpha^2 (4 m_\chi^2 - m_h^2) + s_\alpha^2( 4 m_\chi^2 - m_\phi^2))^2}{v_\phi^2 (4 m_\chi^2 - m_h^2)^2 ( m_\chi^2 - m_\phi^2)^2} \right. \\
 & \hspace{1 in} - \frac{g_p^4}{m_{W_\R}^4 (4 m_\chi^2 - m_{W_\R}^2)^2 ( 2m_\chi^2 - m_{W_\R}^2)^2 (m_\chi^2 + m_{\chi_2}^2 -m_{W_\R})^2} \\
 & \hspace{1 in} \left[ -128 m_\chi^{10} (( 7+ 2\sqrt{2}) m_\chi^4 + 2 (3 + \sqrt{2}) m_\chi^2 m_{\chi_2}^2  + 3 m_{\chi_2}^4)\right. \\
  & \hspace{1 in} + (17 + 12 \sqrt{2}) m_{W_\R}^{14} - 2 m_{W_\R}^{12} (5 ( 4 \sqrt{2}  -9) m_\chi^2 + 2 (8 + 3 \sqrt{2} ) m_{\chi_2}^2) \\
  & \hspace{1 in} - 4 m_{W_\R}^{10} (5 (60 + 17 \sqrt{2}) m_\chi^4 + (50 - 7 \sqrt{2} ) m_\chi^2 m_{\chi_2}^2 - 4 m_{\chi_2}^4)\\
  & \hspace{1 in} + 4 m_{W_\R}^{8} ((997 + 508 \sqrt{2}) m_\chi^6+ 2 ( 229 + 46 \sqrt{2}) m_\chi^4 m_{\chi_2}^2 - 4 m_{\chi_2}^4)\\
  & \hspace{1 in} + 16  m_{W_\R}^6 ((379 + 244\sqrt{2}) m_\chi^8 + 2 ( 125 + 52 \sqrt{2}) m_\chi^6 m_{\chi_2}^2 + 50 m_\chi^4 m_{\chi_2}^4)\\
  & \hspace{1 in}+ 16 m_\chi^6 m_{W_\R}^4 ((249 + 184 \sqrt{2}) m_\chi^4 + 14 (13 + 10 \sqrt{2}) m_\chi^2 m_{\chi_2}^2 + 75 m_{\chi_2}^4)\\
  & \hspace{1 in} \left. \left. - 64 m_\chi^8 m_{W_\R}^2 (7 \sqrt{2} m_\chi^4 + (11 \sqrt{2} -4) m_\chi^2 m_{\chi_2}^2 + 4 m_{\chi_2}^4)\right]\right\}\\
 &\langle \sigma v \rangle_{\chi \chi \to \eta \eta } = \frac{\yc^2 \sqrt{4m_\chi^2 - m_\eta^2}}{8 \pi m_\chi (4m_\chi^2 - m_h^2)^2 (4m_\chi^2 - m_\phi^2)^2} (c_\alpha \xi_{\phi \eta} m_h^2  + 4 m_\chi^2 (\xi_{h \eta} s_\alpha - \xi_{\phi \eta} c_\alpha) - \xi_h s_\alpha m_\phi^2)^2 
\end{align*}
\end{scriptsize}

\subsection{$W^\pm_\R$ DM }
\label{sec:WR}

\begin{scriptsize}
\begin{align*}
&\langle \sigma v \rangle_{W_R W_R \to e_\R e_\R}=\frac{5 g_p^4 v^2}{216 \pi m_{W_R}^2}\\
&\langle \sigma v \rangle_{W_R W_R \to t \bar t}=\frac{3 m_{W_R}  c_\alpha^2 s_\alpha^2 y_t^2 \left(m_h^2-m_\phi^2\right)^2 \left(4 m_{W_R}^2-2 m_t^2\right) \sqrt{4 m_{W_R}^2-m_t^2}}{4 \pi  v_\phi^2 \left(m_h^2-4 m_{W_R}^2\right)^2 \left(m_\phi^2-4 m_{W_R}^2\right)^2}\\
&\langle \sigma v \rangle_{W_R W_R \to Z Z}=\frac{2 m_{W_R} c_\alpha^2  s_\alpha^2 \left(m_h^2-m_\phi^2\right)^2 \sqrt{4 m_{W_R}^2-m_Z^2} \left(4 m_{W_R}^4-4m_{W_R}^2m_Z^2+3 m_Z^4\right)}{3 \pi  v_h^2 v_\phi^2 \left(m_h^2-4 m_{W_R}^2\right)^2 \left(m_\phi^2-4m_{W_R}^2\right)^2}\\
&\langle \sigma v \rangle_{W_R W_R \to W W}=\frac{m_{W_R} c_\alpha^2  s_\alpha^2 \left(m_h^2-m_\phi^2\right)^2 \sqrt{4 m_{W_R}^2-m_W^2} \left(4 m_{W_R}^4-4m_{W_R}^2m_W^2+3 m_W^4\right)}{6 \pi  v_h^2 v_\phi^2 \left(m_h^2-4 m_{W_R}^2\right)^2 \left(m_\phi^2-4m_{W_R}^2\right)^2}\\
&\langle \sigma v \rangle_{W_R W_R \to h h}=\frac{3 m_{W_R}  m_h^4 c_\alpha^2 s_\alpha^2 \left(m_h^2-m_\phi ^2\right)^2}{8 \pi  v_h^2 v_\phi^2 \left(4 m_{W_R} ^2-m_h^2\right)^{3/2} \left(m_\phi^2-4 m_{W_R} ^2\right)^2}\\
&\langle \sigma v \rangle_{W_R W_R \to \phi \phi}=\frac{3 m_{W_R} m_\phi^4 \left(c_\alpha^2 \left(m_h^2-4 m_{W_R}^2\right)+s_\alpha^2 \left(m_\phi^2-4 m_{W_R}^2\right)\right)^2}{8 \pi  v_\phi^4 \left(m_h^2-4 m_{W_R}^2\right)^2 \left(4 m_{W_R}^2-m_\phi^2\right)^{3/2}}\\
&\langle \sigma v \rangle_{W_R W_R \to h \phi}=\frac{3  s_\alpha^2 \sqrt{m_h^4+m_h^2 \left(4 m_{W_R}^2-2 m_\phi^2\right)+\left(m_\phi^2-4 m_{W_R} ^2\right)^2} }{64 \pi  v_h^2 v_\phi^4 \left(m_h^2-4 m_{W_R} ^2\right)^2 \left(m_\phi^2-4 m_{W_R}^2\right)^2}\left(c_\alpha m_\phi^2 v_h \left(m_h^2-4 m_{W_R}^2\right)+m_h^2 s_\alpha v_\phi \left(m_\phi^2-4 m_{W_R}^2\right)\right)^2
  \end{align*}
\end{scriptsize}

\begin{scriptsize}
\begin{align*}
&\langle \sigma v \rangle_{W_R W_R \to \chi \chi}=\frac{\sqrt{4 m_{W_R}^2-m_\chi^2}}{576 \pi  m_{W_R}} [\frac{96 m_{W_R}^2 y_{\chi_1}^2 \left(2 m_{W_R}^2-m_\chi^2\right) \left(c_\alpha^2 \left(m_h^2-4 m_{W_R}^2\right)+s_\alpha^2 \left(m_\phi^2-4 m_{W_R}^2\right)\right)^2}{v_\phi^2 \left(m_h^2-4 m_{W_R}^2\right)^2 \left(m_\phi^2-4 m_{W_R}^2\right)^2}\\
& \hspace{0.8 in} +\frac{g_p^4 \left(4 m_\chi^4+8 m_{W_R}^4-3 m_\chi^2 m_{W_R}^2\right)}{m_{W_R}^2 \left(-m_\chi^2+m_{\chi_2}^2+m_{W_R}^2\right)^2}]\\
&\langle \sigma v \rangle_{W_R W_R \to \eta \eta}=\frac{ m_{W_R} \sqrt{4 m_{W_R}^2-m_\eta^2} \left(c_\alpha \xi_\phi  m_h^2+4 m_{W_R}^2 (\xi_h s_\alpha-c_\alpha \xi_\phi )- \xi_h m_\phi^2 s_\alpha \right)^2}{6 \pi  v_\phi^2 \left(m_h^2-4 m_{W_R}^2\right)^2 \left(m_\phi ^2-4 m_{W_R}^2\right)^2}\\
&\langle \sigma v \rangle_{W_R W_R \to W^3_R W^3_R }=\frac{\sqrt{3} m_{W_R}^6 \left(c_\alpha^2 \left(m_h^2-4 m_{W_R}^2\right)+s_\alpha^2 \left(m\phi^2-4 m_{W_R}^2\right)\right)^2}{2 \pi  v_\phi^4 \left(m_h^2-4 m_{W_R}^2\right)^2 \left(m_\phi^2-4 m_{W_R}^2\right)^2}
\end{align*}
\end{scriptsize}

\subsection{$\eta$ DM }
\label{sec:Eta}
\begin{scriptsize}
\begin{align*}
&\langle \sigma v \rangle_{\eta\eta \to t \bar t}=\frac{9 y_t^2 \left(4 m_\eta^2-2 m_t^2\right) \sqrt{4 m_\eta^2-m_t^2} \left(c_\alpha \xi_h \left(m_\phi^2-4 m_\eta^2\right)+s_\alpha \xi_\phi  \left(m_h^2-4 m_\eta^2\right)\right)^2}{16 \pi  m_\eta^3 \left(m_\phi^2-4 m_\eta^2\right)^2 \left(m_h^2-4 m_\eta^2\right)^2}\\
&\langle \sigma v \rangle_{\eta\eta \to V V}=\frac{  \sqrt{4 m_\eta^2-m_V^2} \left(4 m_\eta^4+3 m_V^4-4 m_\eta^2 m_V^2\right) \left(c_\alpha  \xi_h \left(4 m_\eta^2-m_\phi^2\right)- s_\alpha \xi_\phi  \left(m_h^2-4 m_\eta^2\right)\right)^2}{8 \pi  m_\eta^3 v_h^2 \left(m_\phi^2-4 m_\eta^2\right)^2 \left(m_h^2-4 m_\eta^2\right)^2}\\
&\langle \sigma v \rangle_{\eta\eta \to \chi \chi}=\frac{y^2_{\chi_1} \left(2 m_\eta^2-m_\chi^2\right) \sqrt{4 m_\eta^2-m_\chi^2} \left(4 m_\eta^2 (s_\alpha \xi_h -c_\alpha \xi_\phi )+c_\alpha \xi_\phi  m_h^2-\xi_h m_\phi^2 s_\alpha\right)^2}{8 \pi  m_\eta^3 \left(m_\phi^2-4 m_\eta^2\right)^2 \left(m_h^2-4 m_\eta^2\right)^2}\\
&\langle \sigma v \rangle_{\eta\eta \to W_R W_R}=\frac{9\sqrt{4 m_\eta^2-m_{W_R}^2} \left(4 m_\eta^4+3 m_{W_R}^4-4 m_\eta^2 m_{W_R}^2\right) \left(4 m_\eta^2 (\xi_h s_\alpha-c_\alpha \xi_\phi )+c_\alpha  \xi_\phi  m_h^2-\xi_h m_\phi^2 s_\alpha\right)^2}{8 \pi  m_\eta^3 v_\phi ^2 \left(m_\phi^2-4 m_\eta^2\right)^2 \left(m_h^2-4 m_\eta^2\right)^2}\\
&\langle \sigma v \rangle_{\eta\eta \to h h}=\frac{1}{32 \pi m_\eta^3 v_h^2 \left(m_\phi^2-4 m_\eta^2\right)^2 \left(m_h^2-2 m_\eta^2\right)^2 \left(4 m_\eta^2-m_h^2\right)^{3/2}}\\
& \hspace{0.5 in} \times [ \left(m_h^2-4 m_\eta^2\right) \left(-16m_\eta^4 \xi_h v_h+3 \xi_\phi  m_h^4 s_\alpha+m_\eta^2 \left(4 v_h \left(2 \xi_h^2+2 m_h^2 \xi_h+m_\phi^2 \xi_h\right)-6 \xi_\phi m_h^2 s_\alpha\right)-2 m_\phi^2 v_h \left(\xi_h^2+m_h^2 \xi_h\right)\right)\\
& \hspace{0.5 in} -3 c_\alpha \xi_h m_h^2 \left(4 m_\eta^2-m_\phi^2\right) \left(m_h^2-2 m_\eta^2\right)] ^2\\
&\langle \sigma v \rangle_{\eta\eta \to \phi \phi}=\frac{1}{32 \pi m_\eta^3 v_\phi^2 \left(4 m_\eta^2-m_\phi^2\right)^{3/2} \left(m_\phi^2-2m_\eta^2\right)^2 \left(m_h^2-4 m_\eta^2\right)^2}\\
& \hspace{0.5 in} \times [-8 m_\eta^4 \left(3 m_\phi^2 (-c_\alpha \kappa _\phi +\xi_h s_\alpha+2 \xi_\phi  v_\phi)+2 v_\phi \left(2 \xi_\phi^2+m_h^2 \xi_\phi \right)\right)\\
& \hspace{0.5 in} +2 m_\eta^2 \left(m_\phi^2 \left(-3 c_\alpha \xi_\phi  m_h^2+6 m_h^2 \xi_\phi  v_\phi+4 \xi_\phi^2 v_\phi \right)+m_\phi^4 (-6 c_\alpha \xi_\phi +9 \xi_h s_\alpha+4 \xi_\phi  v_\phi)+4 \xi_\phi ^2 m_h^2 v_\phi \right)\\
& \hspace{0.5 in}+m_h^2 m_\phi^4 (3 c_{\alpha} \xi_\phi -2 \xi_\phi  v_\phi)+64 m_\eta^6 \xi_\phi  v_\phi-2 \xi_\phi ^2 m_h^2 m_\phi^2 v_\phi-3 \xi_h m_\phi^6 s_\alpha]^2\\
&\langle \sigma v \rangle_{\eta\eta \to h \phi}= \frac{\sqrt{\left(m_\phi^2-4 m_\eta^2\right)^2+m_h^4+m_h^2 \left(4 m_\eta^2-2 m_\phi^2\right)} }{256 \pi  m_\eta^4 v_h^2 v_\phi^2 \left(m_\phi ^2-4 m_\eta^2\right)^2 \left(m_h^2-4 m_\eta^2 \right)^2 \left(-4 m_\eta ^2+m_h^2+m_\phi^2\right)^2} \\
& \hspace{0.5 in}\times [4 \xi_\phi  m_\eta^2 v_h \left(4 m_\eta^2-m_\phi^2\right) \left(3 m_\phi^2 s_\alpha-8 \xi_h v_\phi \right)+3 m_h^4 s_\alpha \left(4 \xi_h m_\eta^2 v_\phi+m_\phi^2 (\xi_\phi  v_h-\xi_h v_\phi)\right)\\
&\hspace{0.5 in} -m_h^2 \left(48 \xi_h m_\eta^4 s_\alpha v_\phi-8 m_\eta^2 \left(3 m_\phi^2 s_\alpha (\xi_h v_\phi-\xi_\phi v_h)+4 \xi_h \xi_\phi  v_h v_\phi \right)+3 m_\phi^4 s_\alpha (\xi_h v_\phi -\xi_\phi  v_h)+8 \xi_h \xi_\phi  m_\phi^2 v_h v_\phi \right)]^2
\end{align*}
\end{scriptsize}

\bibliography{DM}

@misc{PGS,
  title = {PGS4},
  howpublished = {\url{http://conway.physics.ucdavis.edu/research/software/pgs/pgs4-general.htm}},
}

@article{Dienes:2011ja,
      author         = "Dienes, Keith R. and Thomas, Brooks",
      title          = "{Dynamical Dark Matter: I. Theoretical Overview}",
      journal        = "Phys. Rev.",
      volume         = "D85",
      year           = "2012",
      pages          = "083523",
      doi            = "10.1103/PhysRevD.85.083523",
      eprint         = "1106.4546",
      archivePrefix  = "arXiv",
      primaryClass   = "hep-ph",
      SLACcitation   = "
}
@inproceedings{Khlopov:1995pa,
      author         = "Khlopov, M. {\relax Yu}.",
      title          = "{Physical arguments, favouring multicomponent dark
                        matter}",
      booktitle      = "{Dark matter in cosmology, clocks and test of fundamental
                        laws. Proceedings, 30th Rencontres de Moriond, 15th
                        Moriond Workshop, Villars sur Ollon, Switzerland, January
                        22-29, 1995}",
      year           = "1995",
      pages          = "133-138",
      SLACcitation   = "
}
@article{Bhattacharya:2017fid,
      author         = "Bhattacharya, Subhaditya and Ghosh, Purusottam and Maity,
                        Tarak Nath and Ray, Tirtha Sankar",
      title          = "{Mitigating Direct Detection Bounds in Non-minimal Higgs
                        Portal Scalar Dark Matter Models}",
      journal        = "JHEP",
      volume         = "10",
      year           = "2017",
      pages          = "088",
      doi            = "10.1007/JHEP10(2017)088",
      eprint         = "1706.04699",
      archivePrefix  = "arXiv",
      primaryClass   = "hep-ph",
      SLACcitation   = "
}
@article{Huang:2015wts,
      author         = "Huang, Wei-Chih and Tsai, Yue-Lin Sming and Yuan,
                        Tzu-Chiang",
      title          = "{G2HDM : Gauged Two Higgs Doublet Model}",
      journal        = "JHEP",
      volume         = "04",
      year           = "2016",
      pages          = "019",
      doi            = "10.1007/JHEP04(2016)019",
      eprint         = "1512.00229",
      archivePrefix  = "arXiv",
      primaryClass   = "hep-ph",
      reportNumber   = "IPMU15-0203, LCTS-2015-41, DO-TH-15-16",
      SLACcitation   = "
}
@article{Dienes:2011sa,
      author         = "Dienes, Keith R. and Thomas, Brooks",
      title          = "{Dynamical Dark Matter: II. An Explicit Model}",
      journal        = "Phys. Rev.",
      volume         = "D85",
      year           = "2012",
      pages          = "083524",
      doi            = "10.1103/PhysRevD.85.083524",
      eprint         = "1107.0721",
      archivePrefix  = "arXiv",
      primaryClass   = "hep-ph",
      SLACcitation   = "
}
@article{Kannike:2012pe,
      author         = "Kannike, Kristjan",
      title          = "{Vacuum Stability Conditions From Copositivity Criteria}",
      journal        = "Eur. Phys. J.",
      volume         = "C72",
      year           = "2012",
      pages          = "2093",
      doi            = "10.1140/epjc/s10052-012-2093-z",
      eprint         = "1205.3781",
      archivePrefix  = "arXiv",
      primaryClass   = "hep-ph",
      SLACcitation   = "
}
@article{Dev:2016qeb,
      author         = "Dev, P. S. Bhupal and Mohapatra, Rabindra N. and Zhang,
                        Yongchao",
      title          = "{Heavy right-handed neutrino dark matter in left-right
                        models}",
      journal        = "Mod. Phys. Lett.",
      volume         = "A32",
      year           = "2017",
      pages          = "1740007",
      doi            = "10.1142/S0217732317400077",
      eprint         = "1610.05738",
      archivePrefix  = "arXiv",
      primaryClass   = "hep-ph",
      reportNumber   = "ULB-TH-16-18",
      SLACcitation   = "
}
@article{Dev:2016xcp,
      author         = "Bhupal Dev, P. S. and Mohapatra, Rabindra N. and Zhang,
                        Yongchao",
      title          = "{Naturally stable right-handed neutrino dark matter}",
      journal        = "JHEP",
      volume         = "11",
      year           = "2016",
      pages          = "077",
      doi            = "10.1007/JHEP11(2016)077",
      eprint         = "1608.06266",
      archivePrefix  = "arXiv",
      primaryClass   = "hep-ph",
      reportNumber   = "ULB-TH-16-14, UMD-PP-016-008",
      SLACcitation   = "
}
@article{Bhattacharya:2016ysw,
      author         = "Bhattacharya, Subhaditya and Poulose, Poulose and Ghosh,
                        Purusottam",
      title          = "{Multipartite Interacting Scalar Dark Matter in the light
                        of updated LUX data}",
      journal        = "JCAP",
      volume         = "1704",
      year           = "2017",
      number         = "04",
      pages          = "043",
      doi            = "10.1088/1475-7516/2017/04/043",
      eprint         = "1607.08461",
      archivePrefix  = "arXiv",
      primaryClass   = "hep-ph",
      SLACcitation   = "
}
@article{Bhattacharya:2018cgx,
      author         = "Bhattacharya, Subhaditya and Ghosh, Purusottam and Sahu,
                        Narendra",
      title          = "{Multipartite Dark Matter with Scalars, Fermions and
                        signatures at LHC}",
      year           = "2018",
      eprint         = "1809.07474",
      archivePrefix  = "arXiv",
      primaryClass   = "hep-ph",
      SLACcitation   = "
}
@article{Cao:2017sju,
      author         = "Cao, Junjie and Feng, Lei and Guo, Xiaofei and Shang,
                        Liangliang and Wang, Fei and Wu, Peiwen and Zu, Lei",
      title          = "{Explaining the DAMPE data with scalar dark matter and
                        gauged $U(1)_{L_e-L_\mu }$ interaction}",
      journal        = "Eur. Phys. J.",
      volume         = "C78",
      year           = "2018",
      number         = "3",
      pages          = "198",
      doi            = "10.1140/epjc/s10052-018-5678-3",
      eprint         = "1712.01244",
      archivePrefix  = "arXiv",
      primaryClass   = "hep-ph",
      SLACcitation   = "
}
@article{Cao:2017rjr,
      author         = "Cao, Junjie and Guo, Xiaofei and Shang, Liangliang and
                        Wang, Fei and Wu, Peiwen and Zu, Lei",
      title          = "{Scalar dark matter explanation of the DAMPE data in the
                        minimal Left-Right symmetric model}",
      journal        = "Phys. Rev.",
      volume         = "D97",
      year           = "2018",
      number         = "6",
      pages          = "063016",
      doi            = "10.1103/PhysRevD.97.063016",
      eprint         = "1712.05351",
      archivePrefix  = "arXiv",
      primaryClass   = "hep-ph",
      SLACcitation   = "
}

@article{Karam:2016rsz,
      author         = "Karam, Alexandros and Tamvakis, Kyriakos",
      title          = "{Dark Matter from a Classically Scale-Invariant
                        $SU(3)_X$}",
      journal        = "Phys. Rev.",
      volume         = "D94",
      year           = "2016",
      number         = "5",
      pages          = "055004",
      doi            = "10.1103/PhysRevD.94.055004",
      eprint         = "1607.01001",
      archivePrefix  = "arXiv",
      primaryClass   = "hep-ph",
      SLACcitation   = "
}
@article{Dienes:2013xff,
      author         = "Dienes, Keith R. and Kumar, Jason and Thomas, Brooks",
      title          = "{Dynamical Dark Matter and the positron excess in light
                        of AMS results}",
      journal        = "Phys. Rev.",
      volume         = "D88",
      year           = "2013",
      number         = "10",
      pages          = "103509",
      doi            = "10.1103/PhysRevD.88.103509",
      eprint         = "1306.2959",
      archivePrefix  = "arXiv",
      primaryClass   = "hep-ph",
      reportNumber   = "UH511-1212-2013",
      SLACcitation   = "
}
@article{Balducci:2018dms,
      author         = "Balducci, Ottavia and Hofmann, Stefan and Kassiteridis,
                        Alexis",
      title          = "{Small-scale structure from charged leptophilia}",
      year           = "2018",
      eprint         = "1812.02182",
      archivePrefix  = "arXiv",
      primaryClass   = "hep-ph",
      SLACcitation   = "
}

@article{2013ApJS..208...19H,
   author = {{Hinshaw}, G. and {Larson}, D. and {Komatsu}, E. and {Spergel}, D.~N. and 
	{Bennett}, C.~L. and {Dunkley}, J. and {Nolta}, M.~R. and {Halpern}, M. and 
	{Hill}, R.~S. and {Odegard}, N. and {Page}, L. and {Smith}, K.~M. and 
	{Weiland}, J.~L. and {Gold}, B. and {Jarosik}, N. and {Kogut}, A. and 
	{Limon}, M. and {Meyer}, S.~S. and {Tucker}, G.~S. and {Wollack}, E. and 
	{Wright}, E.~L.},
    title = "{Nine-year Wilkinson Microwave Anisotropy Probe (WMAP) Observations: Cosmological Parameter Results}",
  journal = {\apjs},
archivePrefix = "arXiv",
   eprint = {1212.5226},
 keywords = {cosmic background radiation, cosmology: observations, dark matter, early universe, instrumentation: detectors, space vehicles, space vehicles: instruments, telescopes },
     year = 2013,
    month = oct,
   volume = 208,
      eid = {19},
    pages = {19},
      doi = {10.1088/0067-0049/208/2/19},
   adsurl = {http://adsabs.harvard.edu/abs/2013ApJS..208...19H},
  adsnote = {Provided by the SAO/NASA Astrophysics Data System}
}
@article{Yao:2018ewe,
      author         = "Yao, Yu-hua and Jin, Chao and Chang, Xiao-chuan",
      title          = "{Test of the 1.4 TeV DAMPE electron excess with
                        preliminary H.E.S.S. measurement}",
      journal        = "Nucl. Phys.",
      volume         = "B934",
      year           = "2018",
      pages          = "396-407",
      doi            = "10.1016/j.nuclphysb.2018.07.018",
      SLACcitation   = "
}
@article{Barger:2007im,
      author         = "Barger, Vernon and Langacker, Paul and McCaskey, Mathew
                        and Ramsey-Musolf, Michael J. and Shaughnessy, Gabe",
      title          = "{LHC Phenomenology of an Extended Standard Model with a
                        Real Scalar Singlet}",
      journal        = "Phys. Rev.",
      volume         = "D77",
      year           = "2008",
      pages          = "035005",
      doi            = "10.1103/PhysRevD.77.035005",
      eprint         = "0706.4311",
      archivePrefix  = "arXiv",
      primaryClass   = "hep-ph",
      reportNumber   = "MAD-PH-07-1492",
      SLACcitation   = "
}
@article{Mohapatra:1979ia,
      author         = "Mohapatra, Rabindra N. and Senjanovic, Goran",
      title          = "{Neutrino Mass and Spontaneous Parity Nonconservation}",
      journal        = "Phys. Rev. Lett.",
      volume         = "44",
      year           = "1980",
      pages          = "912",
      doi            = "10.1103/PhysRevLett.44.912",
      note           = "[,231(1979)]",
      reportNumber   = "MDDP-TR-80-060, MDDP-PP-80-105, CCNY-HEP-79-10",
      SLACcitation   = "
}

@article{Blumenthal:1984bp,
      author         = "Blumenthal, George R. and Faber, S. M. and Primack, Joel
                        R. and Rees, Martin J.",
      title          = "{Formation of Galaxies and Large Scale Structure with
                        Cold Dark Matter}",
      journal        = "Nature",
      volume         = "311",
      year           = "1984",
      pages          = "517-525",
      doi            = "10.1038/311517a0",
      note           = "[,96(1984)]",
      reportNumber   = "SLAC-PUB-3307",
      SLACcitation   = "
}
@article{Tang:2017lfb,
      author         = "Tang, Yi-Lei and Wu, Lei and Zhang, Mengchao and Zheng,
                        Rui",
      title          = "{Lepton-portal Dark Matter in Hidden Valley model and the
                        DAMPE recent results}",
      journal        = "Sci. China Phys. Mech. Astron.",
      volume         = "61",
      year           = "2018",
      number         = "10",
      pages          = "101003",
      doi            = "10.1007/s11433-018-9227-4",
      eprint         = "1711.11058",
      archivePrefix  = "arXiv",
      primaryClass   = "hep-ph",
      SLACcitation   = "
}
@article{Ding:2017jdr,
      author         = "Ding, Ran and Han, Zhi-Long and Feng, Lei and Zhu, Bin",
      title          = "{Confronting the DAMPE Excess with the Scotogenic Type-II
                        Seesaw Model}",
      journal        = "Chin. Phys.",
      volume         = "C42",
      year           = "2018",
      number         = "8",
      pages          = "083104",
      doi            = "10.1088/1674-1137/42/8/083104",
      eprint         = "1712.02021",
      archivePrefix  = "arXiv",
      primaryClass   = "hep-ph",
      SLACcitation   = "
}
@article{Yang:2017cjm,
      author         = "Yang, Fengwei and Su, Meng and Zhao, Yue",
      title          = "{Dark Matter Annihilation from Nearby Ultra-compact Micro
                        Halos to Explain the Tentative Excess at ~1.4 TeV in DAMPE
                        data}",
      year           = "2017",
      eprint         = "1712.01724",
      archivePrefix  = "arXiv",
      primaryClass   = "astro-ph.HE",
      SLACcitation   = "
}
@article{Ghorbani:2017cey,
      author         = "Ghorbani, Karim and Ghorbani, Parsa Hossein",
      title          = "{DAMPE electron-positron excess in leptophilic Z?
                        model}",
      journal        = "JHEP",
      volume         = "05",
      year           = "2018",
      pages          = "125",
      doi            = "10.1007/JHEP05(2018)125",
      eprint         = "1712.01239",
      archivePrefix  = "arXiv",
      primaryClass   = "hep-ph",
      SLACcitation   = "
}
@article{Nomura:2017ohi,
      author         = "Nomura, Takaaki and Okada, Hiroshi",
      title          = "{Radiative seesaw models linking to dark matter
                        candidates inspired by the DAMPE excess}",
      journal        = "Phys. Dark Univ.",
      volume         = "21",
      year           = "2018",
      pages          = "90-95",
      doi            = "10.1016/j.dark.2018.07.003",
      eprint         = "1712.00941",
      archivePrefix  = "arXiv",
      primaryClass   = "hep-ph",
      reportNumber   = "KIAS-P17132",
      SLACcitation   = "
}
@article{Gu:2017lir,
      author         = "Gu, Pei-Hong",
      title          = "{Quasi-degenerate dark matter for DAMPE excess and 3.5
                        keV line}",
      journal        = "Sci. China Phys. Mech. Astron.",
      volume         = "61",
      year           = "2018",
      number         = "10",
      pages          = "101005",
      doi            = "10.1007/s11433-018-9255-x",
      eprint         = "1712.00922",
      archivePrefix  = "arXiv",
      primaryClass   = "hep-ph",
      SLACcitation   = "
}
@article{Li:2017tmd,
      author         = "Li, Tong and Okada, Nobuchika and Shafi, Qaisar",
      title          = "{Scalar dark matter, Type II Seesaw and the DAMPE cosmic
                        ray $e^+ + e^-$ excess}",
      journal        = "Phys. Lett.",
      volume         = "B779",
      year           = "2018",
      pages          = "130-135",
      doi            = "10.1016/j.physletb.2018.02.006",
      eprint         = "1712.00869",
      archivePrefix  = "arXiv",
      primaryClass   = "hep-ph",
      SLACcitation   = "
}
@article{Niu:2017hqe,
      author         = "Niu, Jia-Shu and Li, Tianjun and Ding, Ran and Zhu, Bin
                        and Xue, Hui-Fang and Wang, Yang",
      title          = "{Bayesian analysis of the break in $DAMPE$ lepton
                        spectra}",
      journal        = "Phys. Rev.",
      volume         = "D97",
      year           = "2018",
      number         = "8",
      pages          = "083012",
      doi            = "10.1103/PhysRevD.97.083012",
      eprint         = "1712.00372",
      archivePrefix  = "arXiv",
      primaryClass   = "astro-ph.HE",
      SLACcitation   = "
}
@article{Jin:2017qcv,
      author         = "Jin, Hong-Bo and Yue, Bin and Zhang, Xin and Chen,
                        Xuelei",
      title          = "{Dark matter explanation of the cosmic ray $e^{+} e^{-}$
                        spectrum excess and peak feature observed by the DAMPE
                        experiment}",
      journal        = "Phys. Rev.",
      volume         = "D98",
      year           = "2018",
      number         = "12",
      pages          = "123008",
      doi            = "10.1103/PhysRevD.98.123008",
      eprint         = "1712.00362",
      archivePrefix  = "arXiv",
      primaryClass   = "astro-ph.HE",
      SLACcitation   = "
}
@article{Duan:2017qwj,
      author         = "Duan, Guang Hua and He, Xiao-Gang and Wu, Lei and Yang,
                        Jin Min",
      title          = "{Leptophilic dark matter in gauged $U(1)_{L{_e}-L_{\mu
                        }}$ model in light of DAMPE cosmic ray ${e{^+}} + {e{^-}}$
                        excess}",
      journal        = "Eur. Phys. J.",
      volume         = "C78",
      year           = "2018",
      number         = "4",
      pages          = "323",
      doi            = "10.1140/epjc/s10052-018-5805-1",
      eprint         = "1711.11563",
      archivePrefix  = "arXiv",
      primaryClass   = "hep-ph",
      SLACcitation   = "
}
@article{Gu:2017bdw,
      author         = "Gu, Pei-Hong",
      title          = "{Radiative Dirac neutrino mass, DAMPE dark matter and
                        leptogenesis}",
      year           = "2017",
      eprint         = "1711.11333",
      archivePrefix  = "arXiv",
      primaryClass   = "hep-ph",
      SLACcitation   = "
}
@article{Herrero-Garcia:2018qnz,
      author         = "Herrero-Garcia, Juan and Scaffidi, Andre and White,
                        Martin and Williams, Anthony G.",
      title          = "{On the direct detection of multi-component dark matter:
                        implications of the relic abundance}",
      journal        = "JCAP",
      volume         = "1901",
      year           = "2019",
      number         = "01",
      pages          = "008",
      doi            = "10.1088/1475-7516/2019/01/008",
      eprint         = "1809.06881",
      archivePrefix  = "arXiv",
      primaryClass   = "hep-ph",
      SLACcitation   = "
}
@article{Wang:2018pcc,
      author         = "Wang, BingBing and Bi, XiaoJun and Lin, SuJie and Yin,
                        PengFei",
      title          = "{Explanations of the DAMPE high energy electron/positron
                        spectrum in the dark matter annihilation and pulsar
                        scenarios}",
      journal        = "Sci. China Phys. Mech. Astron.",
      volume         = "61",
      year           = "2018",
      number         = "10",
      pages          = "101004",
      doi            = "10.1007/s11433-018-9244-y",
      SLACcitation   = "
}
@inproceedings{Beck:2018hau,
      author         = "Beck, Geoff and Colafrancesco, Sergio",
      title          = "{Dark matter gets DAMPE}",
      booktitle      = "{61st Annual Conference of the South African Institute of
                        Physics (SAIP2016) Johannesburg, South Africa, July 4-8,
                        2016}",
      year           = "2018",
      eprint         = "1810.07176",
      archivePrefix  = "arXiv",
      primaryClass   = "astro-ph.HE",
      SLACcitation   = "
}
@article{Chao:2017yjg,
      author         = "Chao, Wei and Yuan, Qiang",
      title          = "{The electron-flavored Z'-portal dark matter and the
                        DAMPE cosmic ray excess}",
      year           = "2017",
      eprint         = "1711.11182",
      archivePrefix  = "arXiv",
      primaryClass   = "hep-ph",
      SLACcitation   = "
}
@article{Yao:2018ewe,
      author         = "Yao, Yu-hua and Jin, Chao and Chang, Xiao-chuan",
      title          = "{Test of the 1.4 TeV DAMPE electron excess with
                        preliminary H.E.S.S. measurement}",
      journal        = "Nucl. Phys.",
      volume         = "B934",
      year           = "2018",
      pages          = "396-407",
      doi            = "10.1016/j.nuclphysb.2018.07.018",
      SLACcitation   = "
}
@article{Okada:2017pgr,
      author         = "Okada, Nobuchika and Seto, Osamu",
      title          = "{DAMPE excess from decaying right-handed neutrino dark
                        matter}",
      journal        = "Mod. Phys. Lett.",
      volume         = "A33",
      year           = "2018",
      number         = "27",
      pages          = "1850157",
      doi            = "10.1142/S0217732318501572",
      eprint         = "1712.03652",
      archivePrefix  = "arXiv",
      primaryClass   = "hep-ph",
      reportNumber   = "EPHOU-17-016",
      SLACcitation   = "
}
@article{Liu:2017rgs,
      author         = "Liu, Xuewen and Liu, Zuowei",
      title          = "{TeV dark matter and the DAMPE electron excess}",
      journal        = "Phys. Rev.",
      volume         = "D98",
      year           = "2018",
      number         = "3",
      pages          = "035025",
      doi            = "10.1103/PhysRevD.98.035025",
      eprint         = "1711.11579",
      archivePrefix  = "arXiv",
      primaryClass   = "hep-ph",
      SLACcitation   = "
}

@article{Chakraborty:2012rb,
      author         = "Chakraborty, Indrani and Kundu, Anirban",
      title          = "{Controlling the fine-tuning problem with singlet scalar
                        dark matter}",
      journal        = "Phys. Rev.",
      volume         = "D87",
      year           = "2013",
      number         = "5",
      pages          = "055015",
      doi            = "10.1103/PhysRevD.87.055015",
      eprint         = "1212.0394",
      archivePrefix  = "arXiv",
      primaryClass   = "hep-ph",
      SLACcitation   = "
}
@article{Cao:2017ydw,
      author         = "Cao, Junjie and Feng, Lei and Guo, Xiaofei and Shang,
                        Liangliang and Wang, Fei and Wu, Peiwen",
      title          = "{Scalar dark matter interpretation of the DAMPE data with
                        U(1) gauge interactions}",
      journal        = "Phys. Rev.",
      volume         = "D97",
      year           = "2018",
      number         = "9",
      pages          = "095011",
      doi            = "10.1103/PhysRevD.97.095011",
      eprint         = "1711.11452",
      archivePrefix  = "arXiv",
      primaryClass   = "hep-ph",
      SLACcitation   = "
}
@article{Gu:2017gle,
      author         = "Gu, Pei-Hong and He, Xiao-Gang",
      title          = "{Electrophilic dark matter with dark photon: from DAMPE
                        to direct detection}",
      journal        = "Phys. Lett.",
      volume         = "B778",
      year           = "2018",
      pages          = "292-295",
      doi            = "10.1016/j.physletb.2018.01.057",
      eprint         = "1711.11000",
      archivePrefix  = "arXiv",
      primaryClass   = "hep-ph",
      SLACcitation   = "
}
@article{Duan:2017pkq,
      author         = "Duan, Guang Hua and Feng, Lei and Wang, Fei and Wu, Lei
                        and Yang, Jin Min and Zheng, Rui",
      title          = "{Simplified TeV leptophilic dark matter in light of DAMPE
                        data}",
      journal        = "JHEP",
      volume         = "02",
      year           = "2018",
      pages          = "107",
      doi            = "10.1007/JHEP02(2018)107",
      eprint         = "1711.11012",
      archivePrefix  = "arXiv",
      primaryClass   = "hep-ph",
      SLACcitation   = "
}
@article{Fan:2017sor,
      author         = "Fan, Yi-Zhong and Huang, Wei-Chih and Spinrath, Martin
                        and Tsai, Yue-Lin Sming and Yuan, Qiang",
      title          = "{A model explaining neutrino masses and the DAMPE cosmic
                        ray electron excess}",
      journal        = "Phys. Lett.",
      volume         = "B781",
      year           = "2018",
      pages          = "83-87",
      doi            = "10.1016/j.physletb.2018.03.066",
      eprint         = "1711.10995",
      archivePrefix  = "arXiv",
      primaryClass   = "hep-ph",
      reportNumber   = "CP3-ORIGINS-2017-056, NCTS-PH-1727",
      SLACcitation   = "
}
@article{Weinberg:1979sa,
      author         = "Weinberg, Steven",
      title          = "{Baryon and Lepton Nonconserving Processes}",
      journal        = "Phys. Rev. Lett.",
      volume         = "43",
      year           = "1979",
      pages          = "1566-1570",
      doi            = "10.1103/PhysRevLett.43.1566",
      reportNumber   = "HUTP-79-A050",
      SLACcitation   = "
}
@article{Yuan:2017ysv,
      author         = "Yuan, Qiang and others",
      title          = "{Interpretations of the DAMPE electron data}",
      year           = "2017",
      eprint         = "1711.10989",
      archivePrefix  = "arXiv",
      primaryClass   = "astro-ph.HE",
      SLACcitation   = "
}

@article{Messina:2018fmz,
      author         = "Messina, Marcello",
      title          = "{Latest results of 1 tonne x year Dark Matter Search with
                        XENON1T}",
      booktitle      = "{Proceedings, 2nd World Summit on Exploring the Dark Side
                        of the Universe (EDSU2018): Point a Pitre, Guadeloupe,
                        France, June 25-29, 2018}",
      collaboration  = "XENON",
      journal        = "PoS",
      volume         = "EDSU2018",
      year           = "2018",
      pages          = "017",
      doi            = "10.22323/1.335.0017",
      SLACcitation   = "
}
@article{Tulin:2017ara,
      author         = "Tulin, Sean and Yu, Hai-Bo",
      title          = "{Dark Matter Self-interactions and Small Scale
                        Structure}",
      journal        = "Phys. Rept.",
      volume         = "730",
      year           = "2018",
      pages          = "1-57",
      doi            = "10.1016/j.physrep.2017.11.004",
      eprint         = "1705.02358",
      archivePrefix  = "arXiv",
      primaryClass   = "hep-ph",
      SLACcitation   = "
}
@article{Hambye:2013sna,
      author         = "Hambye, Thomas and Strumia, Alessandro",
      title          = "{Dynamical generation of the weak and Dark Matter scale}",
      journal        = "Phys. Rev.",
      volume         = "D88",
      year           = "2013",
      pages          = "055022",
      doi            = "10.1103/PhysRevD.88.055022",
      eprint         = "1306.2329",
      archivePrefix  = "arXiv",
      primaryClass   = "hep-ph",
      SLACcitation   = "
}
@article{Gabrielli:2013hma,
      author         = "Gabrielli, Emidio and Heikinheimo, Matti and Kannike,
                        Kristjan and Racioppi, Antonio and Raidal, Martti and
                        Spethmann, Christian",
      title          = "{Towards Completing the Standard Model: Vacuum Stability,
                        EWSB and Dark Matter}",
      journal        = "Phys. Rev.",
      volume         = "D89",
      year           = "2014",
      number         = "1",
      pages          = "015017",
      doi            = "10.1103/PhysRevD.89.015017",
      eprint         = "1309.6632",
      archivePrefix  = "arXiv",
      primaryClass   = "hep-ph",
      SLACcitation   = "
}
@article{Baek:2012se,
      author         = "Baek, Seungwon and Ko, P. and Park, Wan-Il and Senaha,
                        Eibun",
      title          = "{Higgs Portal Vector Dark Matter : Revisited}",
      journal        = "JHEP",
      volume         = "05",
      year           = "2013",
      pages          = "036",
      doi            = "10.1007/JHEP05(2013)036",
      eprint         = "1212.2131",
      archivePrefix  = "arXiv",
      primaryClass   = "hep-ph",
      reportNumber   = "KIAS-P12082",
      SLACcitation   = "
}
@article{Drozd:2011aa,
      author         = "Drozd, A. and Grzadkowski, B. and Wudka, Jose",
      title          = "{Multi-Scalar-Singlet Extension of the Standard Model -
                        the Case for Dark Matter and an Invisible Higgs Boson}",
      journal        = "JHEP",
      volume         = "04",
      year           = "2012",
      pages          = "006",
      doi            = "10.1007/JHEP04(2012)006, 10.1007/JHEP11(2014)130",
      note           = "[Erratum: JHEP11,130(2014)]",
      eprint         = "1112.2582",
      archivePrefix  = "arXiv",
      primaryClass   = "hep-ph",
      SLACcitation   = "
}
@article{Gonderinger:2009jp,
      author         = "Gonderinger, Matthew and Li, Yingchuan and Patel, Hiren
                        and Ramsey-Musolf, Michael J.",
      title          = "{Vacuum Stability, Perturbativity, and Scalar Singlet
                        Dark Matter}",
      journal        = "JHEP",
      volume         = "01",
      year           = "2010",
      pages          = "053",
      doi            = "10.1007/JHEP01(2010)053",
      eprint         = "0910.3167",
      archivePrefix  = "arXiv",
      primaryClass   = "hep-ph",
      reportNumber   = "NPAC-09-13",
      SLACcitation   = "
}
@article{Bian:2013wna,
      author         = "Bian, Ligong and Ding, Ran and Zhu, Bin",
      title          = "{Two Component Higgs-Portal Dark Matter}",
      journal        = "Phys. Lett.",
      volume         = "B728",
      year           = "2014",
      pages          = "105-113",
      doi            = "10.1016/j.physletb.2013.11.034",
      eprint         = "1308.3851",
      archivePrefix  = "arXiv",
      primaryClass   = "hep-ph",
      SLACcitation   = "
}
@article{Bennett:2003ba,
      author         = "Bennett, C. L. and others",
      title          = "{The Microwave Anisotropy Probe (MAP) mission}",
      collaboration  = "WMAP",
      journal        = "Astrophys. J.",
      volume         = "583",
      year           = "2003",
      pages          = "1-23",
      doi            = "10.1086/345346",
      eprint         = "astro-ph/0301158",
      archivePrefix  = "arXiv",
      primaryClass   = "astro-ph",
      SLACcitation   = "
}
@article{Steigman:2012nb,
      author         = "Steigman, Gary and Dasgupta, Basudeb and Beacom, John F.",
      title          = "{Precise Relic WIMP Abundance and its Impact on Searches
                        for Dark Matter Annihilation}",
      journal        = "Phys. Rev.",
      volume         = "D86",
      year           = "2012",
      pages          = "023506",
      doi            = "10.1103/PhysRevD.86.023506",
      eprint         = "1204.3622",
      archivePrefix  = "arXiv",
      primaryClass   = "hep-ph",
      SLACcitation   = "
}
@article{Khan:2017ygl,
      author         = "Khan, Najimuddin",
      title          = "{Neutrino mass and the Higgs portal dark matter in the
                        ESSFSM}",
      journal        = "Adv. High Energy Phys.",
      volume         = "2018",
      year           = "2018",
      pages          = "4809682",
      doi            = "10.1155/2018/4809682",
      eprint         = "1707.07300",
      archivePrefix  = "arXiv",
      primaryClass   = "hep-ph",
      SLACcitation   = "
}
@article{YaserAyazi:2018lrv,
      author         = "Yaser Ayazi, Seyed and Mohamadnejad, Ahmad",
      title          = "{Scale-Invariant Two Component Dark Matter}",
      year           = "2018",
      eprint         = "1808.08706",
      archivePrefix  = "arXiv",
      primaryClass   = "hep-ph",
      SLACcitation   = "
}
@article{Chakraborti:2018aae,
      author         = "Chakraborti, Sreemanti and Dutta Banik, Amit and Islam,
                        Rashidul",
      title          = "{Probing Multicomponent Extension of Inert Doublet Model
                        with a Vector Dark Matter}",
      year           = "2018",
      eprint         = "1810.05595",
      archivePrefix  = "arXiv",
      primaryClass   = "hep-ph",
      SLACcitation   = "
}
@article{Bian:2014cja,
      author         = "Bian, Ligong and Li, Tianjun and Shu, Jing and Wang,
                        Xiao-Chuan",
      title          = "{Two component dark matter with multi-Higgs portals}",
      journal        = "JHEP",
      volume         = "03",
      year           = "2015",
      pages          = "126",
      doi            = "10.1007/JHEP03(2015)126",
      eprint         = "1412.5443",
      archivePrefix  = "arXiv",
      primaryClass   = "hep-ph",
      SLACcitation   = "
}
@article{Ahmed:2017dbb,
      author         = "Ahmed, Aqeel and Duch, Mateusz and Grzadkowski, Bohdan
                        and Iglicki, Michal",
      title          = "{Multi-Component Dark Matter: the vector and fermion
                        case}",
      journal        = "Eur. Phys. J.",
      volume         = "C78",
      year           = "2018",
      number         = "11",
      pages          = "905",
      doi            = "10.1140/epjc/s10052-018-6371-2",
      eprint         = "1710.01853",
      archivePrefix  = "arXiv",
      primaryClass   = "hep-ph",
      reportNumber   = "MITP/17-065, MITP-17-065",
      SLACcitation   = "
}
@article{Gross:2015cwa,
      author         = "Gross, Christian and Lebedev, Oleg and Mambrini, Yann",
      title          = "{Non-Abelian gauge fields as dark matter}",
      journal        = "JHEP",
      volume         = "08",
      year           = "2015",
      pages          = "158",
      doi            = "10.1007/JHEP08(2015)158",
      eprint         = "1505.07480",
      archivePrefix  = "arXiv",
      primaryClass   = "hep-ph",
      SLACcitation   = "
}
@article{Barman:2018esi,
      author         = "Barman, Basabendu and Bhattacharya, Subhaditya and
                        Zakeri, Mohammadreza",
      title          = "{Multipartite Dark Matter in $SU(2)_N$ extension of
                        Standard Model and signatures at the LHC}",
      journal        = "JCAP",
      volume         = "1809",
      year           = "2018",
      number         = "09",
      pages          = "023",
      doi            = "10.1088/1475-7516/2018/09/023",
      eprint         = "1806.01129",
      archivePrefix  = "arXiv",
      primaryClass   = "hep-ph",
      SLACcitation   = "
}
@article{Davoudiasl:2013jma,
      author         = "Davoudiasl, Hooman and Lewis, Ian M.",
      title          = "{Dark Matter from Hidden Forces}",
      journal        = "Phys. Rev.",
      volume         = "D89",
      year           = "2014",
      number         = "5",
      pages          = "055026",
      doi            = "10.1103/PhysRevD.89.055026",
      eprint         = "1309.6640",
      archivePrefix  = "arXiv",
      primaryClass   = "hep-ph",
      SLACcitation   = "
}
@inproceedings{Yamanaka:2015tba,
      author         = "Yamanaka, Nodoka and Fujibayashi, Sho and Gongyo, Shinya
                        and Iida, Hideaki",
      title          = "{Dark Matter in the Nonabelian Hidden Gauge Theory}",
      booktitle      = "{2nd Toyama International Workshop on Higgs as a Probe of
                        New Physics (HPNP2015) Toyama, Japan, February 11-15,
                        2015}",
      year           = "2015",
      eprint         = "1504.08121",
      archivePrefix  = "arXiv",
      primaryClass   = "hep-ph",
      SLACcitation   = "
}
@article{Biswas:2015sva,
      author         = "Biswas, Anirban and Majumdar, Debasish and Roy, Probir",
      title          = "{Nonthermal two component dark matter model for Fermi-LAT
                        ?-ray excess and 3.55 keV X-ray line}",
      journal        = "JHEP",
      volume         = "04",
      year           = "2015",
      pages          = "065",
      doi            = "10.1007/JHEP04(2015)065",
      eprint         = "1501.02666",
      archivePrefix  = "arXiv",
      primaryClass   = "hep-ph",
      SLACcitation   = "
}
@article{DiFranzo:2016uzc,
      author         = "DiFranzo, Anthony and Mohlabeng, Gopolang",
      title          = "{Multi-component Dark Matter through a Radiative Higgs
                        Portal}",
      journal        = "JHEP",
      volume         = "01",
      year           = "2017",
      pages          = "080",
      doi            = "10.1007/JHEP01(2017)080",
      eprint         = "1610.07606",
      archivePrefix  = "arXiv",
      primaryClass   = "hep-ph",
      reportNumber   = "FERMILAB-PUB-16-456-T, UCI-HEP-TR-2016-19",
      SLACcitation   = "
}
@article{DuttaBanik:2016jzv,
      author         = "Dutta Banik, Amit and Pandey, Madhurima and Majumdar,
                        Debasish and Biswas, Anirban",
      title          = "{Two component WIMP?FImP dark matter model with singlet
                        fermion, scalar and pseudo scalar}",
      journal        = "Eur. Phys. J.",
      volume         = "C77",
      year           = "2017",
      number         = "10",
      pages          = "657",
      doi            = "10.1140/epjc/s10052-017-5221-y",
      eprint         = "1612.08621",
      archivePrefix  = "arXiv",
      primaryClass   = "hep-ph",
      SLACcitation   = "
}
@article{Profumo:2014opa,
      author         = "Profumo, Stefano and Ramsey-Musolf, Michael J. and
                        Wainwright, Carroll L. and Winslow, Peter",
      title          = "{Singlet-catalyzed electroweak phase transitions and
                        precision Higgs boson studies}",
      journal        = "Phys. Rev.",
      volume         = "D91",
      year           = "2015",
      number         = "3",
      pages          = "035018",
      doi            = "10.1103/PhysRevD.91.035018",
      eprint         = "1407.5342",
      archivePrefix  = "arXiv",
      primaryClass   = "hep-ph",
      SLACcitation   = "
}
@article{Noble:2007kk,
      author         = "Noble, Andrew and Perelstein, Maxim",
      title          = "{Higgs self-coupling as a probe of electroweak phase
                        transition}",
      journal        = "Phys. Rev.",
      volume         = "D78",
      year           = "2008",
      pages          = "063518",
      doi            = "10.1103/PhysRevD.78.063518",
      eprint         = "0711.3018",
      archivePrefix  = "arXiv",
      primaryClass   = "hep-ph",
      SLACcitation   = "
}
@article{Damgaard:2013kva,
      author         = "Damgaard, Poul H. and O'Connell, Donal and Petersen,
                        Troels C. and Tranberg, Anders",
      title          = "{Constraints on New Physics from Baryogenesis and Large
                        Hadron Collider Data}",
      journal        = "Phys. Rev. Lett.",
      volume         = "111",
      year           = "2013",
      number         = "22",
      pages          = "221804",
      doi            = "10.1103/PhysRevLett.111.221804",
      eprint         = "1305.4362",
      archivePrefix  = "arXiv",
      primaryClass   = "hep-ph",
      SLACcitation   = "
}
@article{Farina:2013ssa,
      author         = "Farina, Marco and Perelstein, Maxim and Rey-Le Lorier,
                        Nicolas",
      title          = "{Higgs Couplings and Naturalness}",
      journal        = "Phys. Rev.",
      volume         = "D90",
      year           = "2014",
      number         = "1",
      pages          = "015014",
      doi            = "10.1103/PhysRevD.90.015014",
      eprint         = "1305.6068",
      archivePrefix  = "arXiv",
      primaryClass   = "hep-ph",
      SLACcitation   = "
}
@article{Craig:2013xia,
      author         = "Craig, Nathaniel and Englert, Christoph and McCullough,
                        Matthew",
      title          = "{New Probe of Naturalness}",
      journal        = "Phys. Rev. Lett.",
      volume         = "111",
      year           = "2013",
      number         = "12",
      pages          = "121803",
      doi            = "10.1103/PhysRevLett.111.121803",
      eprint         = "1305.5251",
      archivePrefix  = "arXiv",
      primaryClass   = "hep-ph",
      reportNumber   = "IPPP-13-30, DCPT-13-60, MIT-CTP-4462, RU-NHETC-2013-12",
      SLACcitation   = "
}
@article{Antipin:2013exa,
      author         = "Antipin, Oleg and Mojaza, Matin and Sannino, Francesco",
      title          = "{Conformal Extensions of the Standard Model with Veltman
                        Conditions}",
      journal        = "Phys. Rev.",
      volume         = "D89",
      year           = "2014",
      number         = "8",
      pages          = "085015",
      doi            = "10.1103/PhysRevD.89.085015",
      eprint         = "1310.0957",
      archivePrefix  = "arXiv",
      primaryClass   = "hep-ph",
      reportNumber   = "CP3-ORIGINS-2013-039, DIAS-2013-39",
      SLACcitation   = "
}
@article{Karahan:2014ola,
      author         = "Karahan, Canan Nurhan and Korutlu, Beste",
      title          = "{Effects of a Real Singlet Scalar on Veltman Condition}",
      journal        = "Phys. Lett.",
      volume         = "B732",
      year           = "2014",
      pages          = "320-324",
      doi            = "10.1016/j.physletb.2014.03.063",
      eprint         = "1404.0175",
      archivePrefix  = "arXiv",
      primaryClass   = "hep-ph",
      reportNumber   = "IZTECH-PHYS-2014-05",
      SLACcitation   = "
}
@article{Grzadkowski:2009mj,
      author         = "Grzadkowski, Bohdan and Wudka, Jose",
      title          = "{Pragmatic approach to the little hierarchy problem: the
                        case for Dark Matter and neutrino physics}",
      journal        = "Phys. Rev. Lett.",
      volume         = "103",
      year           = "2009",
      pages          = "091802",
      doi            = "10.1103/PhysRevLett.103.091802",
      eprint         = "0902.0628",
      archivePrefix  = "arXiv",
      primaryClass   = "hep-ph",
      reportNumber   = "IFT-09-01, UCRHEP-T463",
      SLACcitation   = "
}
@article{Esch:2014jpa,
      author         = "Esch, Sonja and Klasen, Michael and Yaguna, Carlos E.",
      title          = "{A minimal model for two-component dark matter}",
      journal        = "JHEP",
      volume         = "09",
      year           = "2014",
      pages          = "108",
      doi            = "10.1007/JHEP09(2014)108",
      eprint         = "1406.0617",
      archivePrefix  = "arXiv",
      primaryClass   = "hep-ph",
      reportNumber   = "MS-TP-14-22",
      SLACcitation   = "
}

@article{Kajiyama:2013rla,
      author         = "Kajiyama, Yuji and Okada, Hiroshi and Toma, Takashi",
      title          = "{Multicomponent dark matter particles in a two-loop
                        neutrino model}",
      journal        = "Phys. Rev.",
      volume         = "D88",
      year           = "2013",
      number         = "1",
      pages          = "015029",
      doi            = "10.1103/PhysRevD.88.015029",
      eprint         = "1303.7356",
      archivePrefix  = "arXiv",
      primaryClass   = "hep-ph",
      reportNumber   = "KIAS-P13014, IPPP-13-14, DCPT-13-28",
      SLACcitation   = "
}
@article{Bhattacharya:2013hva,
      author         = "Bhattacharya, Subhaditya and Drozd, Aleksandra and
                        Grzadkowski, Bohdan and Wudka, Jose",
      title          = "{Two-Component Dark Matter}",
      journal        = "JHEP",
      volume         = "10",
      year           = "2013",
      pages          = "158",
      doi            = "10.1007/JHEP10(2013)158",
      eprint         = "1309.2986",
      archivePrefix  = "arXiv",
      primaryClass   = "hep-ph",
      reportNumber   = "UCRHEP-T537",
      SLACcitation   = "
}
@article{Chialva:2012rq,
      author         = "Chialva, Diego and Dev, P. S. Bhupal and Mazumdar,
                        Anupam",
      title          = "{Multiple dark matter scenarios from ubiquitous stringy
                        throats}",
      journal        = "Phys. Rev.",
      volume         = "D87",
      year           = "2013",
      number         = "6",
      pages          = "063522",
      doi            = "10.1103/PhysRevD.87.063522",
      eprint         = "1211.0250",
      archivePrefix  = "arXiv",
      primaryClass   = "hep-ph",
      reportNumber   = "MAN-HEP-2012-016",
      SLACcitation   = "
}

@article{Aoki:2012ub,
      author         = "Aoki, Mayumi and Duerr, Michael and Kubo, Jisuke and
                        Takano, Hiroshi",
      title          = "{Multi-Component Dark Matter Systems and Their
                        Observation Prospects}",
      journal        = "Phys. Rev.",
      volume         = "D86",
      year           = "2012",
      pages          = "076015",
      doi            = "10.1103/PhysRevD.86.076015",
      eprint         = "1207.3318",
      archivePrefix  = "arXiv",
      primaryClass   = "hep-ph",
      reportNumber   = "KANAZAWA-12-07",
      SLACcitation   = "
}
@article{Baer:2011hx,
      author         = "Baer, Howard and Lessa, Andre and Rajagopalan, Shibi and
                        Sreethawong, Warintorn",
      title          = "{Mixed axion/neutralino cold dark matter in
                        supersymmetric models}",
      journal        = "JCAP",
      volume         = "1106",
      year           = "2011",
      pages          = "031",
      doi            = "10.1088/1475-7516/2011/06/031",
      eprint         = "1103.5413",
      archivePrefix  = "arXiv",
      primaryClass   = "hep-ph",
      SLACcitation   = "
}
@article{Feldman:2010wy,
      author         = "Feldman, Daniel and Liu, Zuowei and Nath, Pran and Peim,
                        Gregory",
      title          = "{Multicomponent Dark Matter in Supersymmetric Hidden
                        Sector Extensions}",
      journal        = "Phys. Rev.",
      volume         = "D81",
      year           = "2010",
      pages          = "095017",
      doi            = "10.1103/PhysRevD.81.095017",
      eprint         = "1004.0649",
      archivePrefix  = "arXiv",
      primaryClass   = "hep-ph",
      reportNumber   = "MCTP-10-15, YITP-SB-10-08, NUB-3266",
      SLACcitation   = "
}
@article{Gao:2010pg,
      author         = "Gao, Xin and Kang, Zhaofeng and Li, Tianjun",
      title          = "{The Supersymmetric Standard Models with Decay and Stable
                        Dark Matters}",
      journal        = "Eur. Phys. J.",
      volume         = "C69",
      year           = "2010",
      pages          = "467-480",
      doi            = "10.1140/epjc/s10052-010-1418-z",
      eprint         = "1001.3278",
      archivePrefix  = "arXiv",
      primaryClass   = "hep-ph",
      SLACcitation   = "
}
@article{Profumo:2009tb,
      author         = "Profumo, Stefano and Sigurdson, Kris and Ubaldi, Lorenzo",
      title          = "{Can we discover multi-component WIMP dark matter?}",
      journal        = "JCAP",
      volume         = "0912",
      year           = "2009",
      pages          = "016",
      doi            = "10.1088/1475-7516/2009/12/016",
      eprint         = "0907.4374",
      archivePrefix  = "arXiv",
      primaryClass   = "hep-ph",
      SLACcitation   = "
}
@article{Duda:2002hf,
      author         = "Duda, Gintaras and Gelmini, Graciela and Gondolo, Paolo
                        and Edsjo, Joakim and Silk, Joseph",
      title          = "{Indirect detection of a subdominant density component of
                        cold dark matter}",
      journal        = "Phys. Rev.",
      volume         = "D67",
      year           = "2003",
      pages          = "023505",
      doi            = "10.1103/PhysRevD.67.023505",
      eprint         = "hep-ph/0209266",
      archivePrefix  = "arXiv",
      primaryClass   = "hep-ph",
      reportNumber   = "UCLA-02-TEP-22, CWRU-P11-02, NSF-ITP-02-96",
      SLACcitation   = "
}
@article{Duda:2001ae,
      author         = "Duda, Gintaras and Gelmini, Graciela and Gondolo, Paolo",
      title          = "{Detection of a subdominant density component of cold
                        dark matter}",
      journal        = "Phys. Lett.",
      volume         = "B529",
      year           = "2002",
      pages          = "187-192",
      doi            = "10.1016/S0370-2693(02)01266-2",
      eprint         = "hep-ph/0102200",
      archivePrefix  = "arXiv",
      primaryClass   = "hep-ph",
      reportNumber   = "UCLA-01-TEP-3, CWRU-P4-01",
      SLACcitation   = "
}
@article{Chao:2017emq,
      author         = "Chao, Wei and Guo, Huai-Ke and Li, Hao-Lin and Shu, Jing",
      title          = "{Electron Flavored Dark Matter}",
      journal        = "Phys. Lett.",
      volume         = "B782",
      year           = "2018",
      pages          = "517-522",
      doi            = "10.1016/j.physletb.2018.05.075",
      eprint         = "1712.00037",
      archivePrefix  = "arXiv",
      primaryClass   = "hep-ph",
      SLACcitation   = "
}

@article{Ge:2017tkd,
      author         = "Ge, Shao-Feng and He, Hong-Jian and Wang, Yu-Chen",
      title          = "{Flavor Structure of the Cosmic-Ray Electron/Positron
                        Excesses at DAMPE}",
      journal        = "Phys. Lett.",
      volume         = "B781",
      year           = "2018",
      pages          = "88-94",
      doi            = "10.1016/j.physletb.2018.03.065",
      eprint         = "1712.02744",
      archivePrefix  = "arXiv",
      primaryClass   = "astro-ph.HE",
      reportNumber   = "IPMU-0173",
      SLACcitation   = "
}

@article{Nomura:2018jkd,
      author         = "Nomura, Takaaki and Okada, Hiroshi and Wu, Peiwen",
      title          = "{A radiative neutrino mass model in light of DAMPE excess
                        with hidden gauged $U(1)$ symmetry}",
      journal        = "JCAP",
      volume         = "1805",
      year           = "2018",
      number         = "05",
      pages          = "053",
      doi            = "10.1088/1475-7516/2018/05/053",
      eprint         = "1801.04729",
      archivePrefix  = "arXiv",
      primaryClass   = "hep-ph",
      SLACcitation   = "
}
@article{Niu:2017lts,
      author         = "Niu, Jia-Shu and Li, Tianjun and Xu, Fang-Zhou",
      title          = "{A Simple and Natural Interpretations of the DAMPE Cosmic
                        Ray Electron/Positron Spectrum within Two Sigma
                        Deviations}",
      year           = "2017",
      eprint         = "1712.09586",
      archivePrefix  = "arXiv",
      primaryClass   = "hep-ph",
      SLACcitation   = "
}
@article{Fowlie:2017fya,
      author         = "Fowlie, Andrew",
      title          = "{DAMPE squib? Significance of the 1.4 TeV DAMPE excess}",
      journal        = "Phys. Lett.",
      volume         = "B780",
      year           = "2018",
      pages          = "181-184",
      doi            = "10.1016/j.physletb.2018.03.006",
      eprint         = "1712.05089",
      archivePrefix  = "arXiv",
      primaryClass   = "hep-ph",
      reportNumber   = "COEPP-MN-17-22",
      SLACcitation   = "
}

@article{Huang:2017egk,
      author         = "Huang, Xian-Jun and Wu, Yue-Liang and Zhang, Wei-Hong and
                        Zhou, Yu-Feng",
      title          = "{Origins of sharp cosmic-ray electron structures and the
                        DAMPE excess}",
      journal        = "Phys. Rev.",
      volume         = "D97",
      year           = "2018",
      number         = "9",
      pages          = "091701",
      doi            = "10.1103/PhysRevD.97.091701",
      eprint         = "1712.00005",
      archivePrefix  = "arXiv",
      primaryClass   = "astro-ph.HE",
      SLACcitation   = "
}


@article{Ade:2015xua,
      author         = "Ade, P. A. R. and others",
      title          = "{Planck 2015 results. XIII. Cosmological parameters}",
      collaboration  = "Planck",
      journal        = "Astron. Astrophys.",
      volume         = "594",
      year           = "2016",
      pages          = "A13",
      doi            = "10.1051/0004-6361/201525830",
      eprint         = "1502.01589",
      archivePrefix  = "arXiv",
      primaryClass   = "astro-ph.CO",
      SLACcitation   = "
}


@article{Akerib:2016vxi,
      author         = "Akerib, D. S. and others",
      title          = "{Results from a search for dark matter in the complete
                        LUX exposure}",
      collaboration  = "LUX",
      journal        = "Phys. Rev. Lett.",
      volume         = "118",
      year           = "2017",
      number         = "2",
      pages          = "021303",
      doi            = "10.1103/PhysRevLett.118.021303",
      eprint         = "1608.07648",
      archivePrefix  = "arXiv",
      primaryClass   = "astro-ph.CO",
      SLACcitation   = "
}

@article{Aprile:2012nq,
      author         = "Aprile, E. and others",
      title          = "{Dark Matter Results from 225 Live Days of XENON100
                        Data}",
      collaboration  = "XENON100",
      journal        = "Phys. Rev. Lett.",
      volume         = "109",
      year           = "2012",
      pages          = "181301",
      doi            = "10.1103/PhysRevLett.109.181301",
      eprint         = "1207.5988",
      archivePrefix  = "arXiv",
      primaryClass   = "astro-ph.CO",
      SLACcitation   = "
}



@article{Abdollahi:2017nat,
      author         = "Abdollahi, S. and others",
      title          = "{Cosmic-ray electron-positron spectrum from 7 GeV to
                        2 TeV with the Fermi Large Area Telescope}",
      collaboration  = "Fermi-LAT",
      journal        = "Phys. Rev.",
      volume         = "D95",
      year           = "2017",
      number         = "8",
      pages          = "082007",
      doi            = "10.1103/PhysRevD.95.082007",
      eprint         = "1704.07195",
      archivePrefix  = "arXiv",
      primaryClass   = "astro-ph.HE",
      reportNumber   = "FERMILAB-PUB-17-188-AE",
      SLACcitation   = "
}

@article{Aguilar:2016vqr,
      author         = "Aguilar, M. and others",
      title          = "{Precision Measurement of the Boron to Carbon Flux Ratio
                        in Cosmic Rays from 1.9 GV to 2.6 TV with the Alpha
                        Magnetic Spectrometer on the International Space Station}",
      collaboration  = "AMS",
      journal        = "Phys. Rev. Lett.",
      volume         = "117",
      year           = "2016",
      number         = "23",
      pages          = "231102",
      doi            = "10.1103/PhysRevLett.117.231102",
      SLACcitation   = "
}


@article{Adriani:2017efm,
      author         = "Adriani, O. and others",
      title          = "{Energy Spectrum of Cosmic-Ray Electron and Positron from
                        10 GeV to 3 TeV Observed with the Calorimetric Electron
                        Telescope on the International Space Station}",
      collaboration  = "CALET",
      journal        = "Phys. Rev. Lett.",
      volume         = "119",
      year           = "2017",
      number         = "18",
      pages          = "181101",
      doi            = "10.1103/PhysRevLett.119.181101",
      eprint         = "1712.01711",
      archivePrefix  = "arXiv",
      primaryClass   = "astro-ph.HE",
      SLACcitation   = "
}

@article{Ambrosi:2017wek,
      author         = "Ambrosi, G. and others",
      title          = "{Direct detection of a break in the teraelectronvolt
                        cosmic-ray spectrum of electrons and positrons}",
      collaboration  = "DAMPE",
      journal        = "Nature",
      volume         = "552",
      year           = "2017",
      pages          = "63-66",
      doi            = "10.1038/nature24475",
      eprint         = "1711.10981",
      archivePrefix  = "arXiv",
      primaryClass   = "astro-ph.HE",
      SLACcitation   = "
}

@article{Zhao:2017nrt,
      author         = "Zhao, Yue and Fang, Ke and Su, Meng and Miller, M.
                        Coleman",
      title          = "{A Strong Test of the Dark Matter Origin of the 1.4 TeV
                        DAMPE Signal Using IceCube Neutrinos}",
      year           = "2017",
      eprint         = "1712.03210",
      archivePrefix  = "arXiv",
      primaryClass   = "astro-ph.HE",
      SLACcitation   = "
}
@article{Ackermann:2013yva,
      author         = "Ackermann, M. and others",
      title          = "{Dark matter constraints from observations of 25
                        Milky Way satellite galaxies with the Fermi Large Area
                        Telescope}",
      collaboration  = "Fermi-LAT",
      journal        = "Phys. Rev.",
      volume         = "D89",
      year           = "2014",
      pages          = "042001",
      doi            = "10.1103/PhysRevD.89.042001",
      eprint         = "1310.0828",
      archivePrefix  = "arXiv",
      primaryClass   = "astro-ph.HE",
      reportNumber   = "FERMILAB-PUB-13-683-A",
      SLACcitation   = "
}
@article{Ade:2015xua,
      author         = "Ade, P. A. R. and others",
      title          = "{Planck 2015 results. XIII. Cosmological parameters}",
      collaboration  = "Planck",
      journal        = "Astron. Astrophys.",
      volume         = "594",
      year           = "2016",
      pages          = "A13",
      doi            = "10.1051/0004-6361/201525830",
      eprint         = "1502.01589",
      archivePrefix  = "arXiv",
      primaryClass   = "astro-ph.CO",
      SLACcitation   = "
}

@article{Springel:2005nw,
      author         = "Springel, Volker and others",
      title          = "{Simulating the joint evolution of quasars, galaxies and
                        their large-scale distribution}",
      journal        = "Nature",
      volume         = "435",
      year           = "2005",
      pages          = "629-636",
      doi            = "10.1038/nature03597",
      eprint         = "astro-ph/0504097",
      archivePrefix  = "arXiv",
      primaryClass   = "astro-ph",
      SLACcitation   = "
}

@article{Baek:2012se,
      author         = "Baek, Seungwon and Ko, P. and Park, Wan-Il and Senaha,
                        Eibun",
      title          = "{Higgs Portal Vector Dark Matter : Revisited}",
      journal        = "JHEP",
      volume         = "05",
      year           = "2013",
      pages          = "036",
      doi            = "10.1007/JHEP05(2013)036",
      eprint         = "1212.2131",
      archivePrefix  = "arXiv",
      primaryClass   = "hep-ph",
      reportNumber   = "KIAS-P12082",
      SLACcitation   = "
}

@article{Casas:2017jjg,
      author         = "Casas, J. Alberto and Cerdeño, David G. and Moreno,
                        Jesus M. and Quilis, Javier",
      title          = "{Reopening the Higgs portal for single scalar dark
                        matter}",
      journal        = "JHEP",
      volume         = "05",
      year           = "2017",
      pages          = "036",
      doi            = "10.1007/JHEP05(2017)036",
      eprint         = "1701.08134",
      archivePrefix  = "arXiv",
      primaryClass   = "hep-ph",
      reportNumber   = "IFT-UAM-CSIC-16-113, IPPP-16-107, DCTP-16-214",
      SLACcitation   = "
}
@article{Belanger:2008sj,
      author         = "Belanger, G. and Boudjema, F. and Pukhov, A. and Semenov,
                        A.",
      title          = "{Dark matter direct detection rate in a generic model
                        with micrOMEGAs 2.2}",
      journal        = "Comput. Phys. Commun.",
      volume         = "180",
      year           = "2009",
      pages          = "747-767",
      doi            = "10.1016/j.cpc.2008.11.019",
      eprint         = "0803.2360",
      archivePrefix  = "arXiv",
      primaryClass   = "hep-ph",
      reportNumber   = "LAPTH-1237-08",
      SLACcitation   = "
}
@article{Djouadi:2012zc,
      author         = "Djouadi, Abdelhak and Falkowski, Adam and Mambrini, Yann
                        and Quevillon, Jeremie",
      title          = "{Direct Detection of Higgs-Portal Dark Matter at the
                        LHC}",
      journal        = "Eur. Phys. J.",
      volume         = "C73",
      year           = "2013",
      number         = "6",
      pages          = "2455",
      doi            = "10.1140/epjc/s10052-013-2455-1",
      eprint         = "1205.3169",
      archivePrefix  = "arXiv",
      primaryClass   = "hep-ph",
      reportNumber   = "LPT-ORSAY-12-42",
      SLACcitation   = "
}
@article{Giudice:2012ms,
      author         = "Giudice, G. F. and Paradisi, P. and Passera, M.",
      title          = "{Testing new physics with the electron g-2}",
      journal        = "JHEP",
      volume         = "11",
      year           = "2012",
      pages          = "113",
      doi            = "10.1007/JHEP11(2012)113",
      eprint         = "1208.6583",
      archivePrefix  = "arXiv",
      primaryClass   = "hep-ph",
      reportNumber   = "CERN-PH-TH-2012-017-a",
      SLACcitation   = "
}
@article{Jegerlehner:2009ry,
      author         = "Jegerlehner, Fred and Nyffeler, Andreas",
      title          = "{The Muon g-2}",
      journal        = "Phys. Rept.",
      volume         = "477",
      year           = "2009",
      pages          = "1-110",
      doi            = "10.1016/j.physrep.2009.04.003",
      eprint         = "0902.3360",
      archivePrefix  = "arXiv",
      primaryClass   = "hep-ph",
      reportNumber   = "HU-EP-09-07, HRI-P-09-02-001, RECAPP-HRI-2009-003",
      SLACcitation   = "
}
@article{DiChiara:2017cjq,
      author         = "Di Chiara, Stefano and Fowlie, Andrew and Fraser, Sean
                        and Marzo, Carlo and Marzola, Luca and Raidal, Martti and
                        Spethmann, Christian",
      title          = "{Minimal flavor-changing $Z'$ models and muon $g-2$ after
                        the $R_{K^*}$ measurement}",
      journal        = "Nucl. Phys.",
      volume         = "B923",
      year           = "2017",
      pages          = "245-257",
      doi            = "10.1016/j.nuclphysb.2017.08.003",
      eprint         = "1704.06200",
      archivePrefix  = "arXiv",
      primaryClass   = "hep-ph",
      SLACcitation   = "
}

\providecommand{\href}[2]{#2}\begingroup\raggedright\begin{thebibliography}{10}

\bibitem{Dienes:2011ja}
K.~R. Dienes and B.~Thomas, {\it {Dynamical Dark Matter: I. Theoretical
  Overview}},  {\em Phys. Rev.} {\bf D85} (2012) 083523,
  [\href{http://xxx.lanl.gov/abs/1106.4546}{{\tt 1106.4546}}].

\bibitem{Dienes:2011sa}
K.~R. Dienes and B.~Thomas, {\it {Dynamical Dark Matter: II. An Explicit
  Model}},  {\em Phys. Rev.} {\bf D85} (2012) 083524,
  [\href{http://xxx.lanl.gov/abs/1107.0721}{{\tt 1107.0721}}].

\bibitem{Bian:2013wna}
L.~Bian, R.~Ding, and B.~Zhu, {\it {Two Component Higgs-Portal Dark Matter}},
  {\em Phys. Lett.} {\bf B728} (2014) 105--113,
  [\href{http://xxx.lanl.gov/abs/1308.3851}{{\tt 1308.3851}}].

\bibitem{Duda:2001ae}
G.~Duda, G.~Gelmini, and P.~Gondolo, {\it {Detection of a subdominant density
  component of cold dark matter}},  {\em Phys. Lett.} {\bf B529} (2002)
  187--192, [\href{http://xxx.lanl.gov/abs/hep-ph/0102200}{{\tt
  hep-ph/0102200}}].

\bibitem{Duda:2002hf}
G.~Duda, G.~Gelmini, P.~Gondolo, J.~Edsjo, and J.~Silk, {\it {Indirect
  detection of a subdominant density component of cold dark matter}},  {\em
  Phys. Rev.} {\bf D67} (2003) 023505,
  [\href{http://xxx.lanl.gov/abs/hep-ph/0209266}{{\tt hep-ph/0209266}}].

\bibitem{Profumo:2009tb}
S.~Profumo, K.~Sigurdson, and L.~Ubaldi, {\it {Can we discover multi-component
  WIMP dark matter?}},  {\em JCAP} {\bf 0912} (2009) 016,
  [\href{http://xxx.lanl.gov/abs/0907.4374}{{\tt 0907.4374}}].

\bibitem{Gao:2010pg}
X.~Gao, Z.~Kang, and T.~Li, {\it {The Supersymmetric Standard Models with Decay
  and Stable Dark Matters}},  {\em Eur. Phys. J.} {\bf C69} (2010) 467--480,
  [\href{http://xxx.lanl.gov/abs/1001.3278}{{\tt 1001.3278}}].

\bibitem{Feldman:2010wy}
D.~Feldman, Z.~Liu, P.~Nath, and G.~Peim, {\it {Multicomponent Dark Matter in
  Supersymmetric Hidden Sector Extensions}},  {\em Phys. Rev.} {\bf D81} (2010)
  095017, [\href{http://xxx.lanl.gov/abs/1004.0649}{{\tt 1004.0649}}].

\bibitem{Baer:2011hx}
H.~Baer, A.~Lessa, S.~Rajagopalan, and W.~Sreethawong, {\it {Mixed
  axion/neutralino cold dark matter in supersymmetric models}},  {\em JCAP}
  {\bf 1106} (2011) 031, [\href{http://xxx.lanl.gov/abs/1103.5413}{{\tt
  1103.5413}}].

\bibitem{Aoki:2012ub}
M.~Aoki, M.~Duerr, J.~Kubo, and H.~Takano, {\it {Multi-Component Dark Matter
  Systems and Their Observation Prospects}},  {\em Phys. Rev.} {\bf D86} (2012)
  076015, [\href{http://xxx.lanl.gov/abs/1207.3318}{{\tt 1207.3318}}].

\bibitem{Chialva:2012rq}
D.~Chialva, P.~S.~B. Dev, and A.~Mazumdar, {\it {Multiple dark matter scenarios
  from ubiquitous stringy throats}},  {\em Phys. Rev.} {\bf D87} (2013), no.~6
  063522, [\href{http://xxx.lanl.gov/abs/1211.0250}{{\tt 1211.0250}}].

\bibitem{Bhattacharya:2013hva}
S.~Bhattacharya, A.~Drozd, B.~Grzadkowski, and J.~Wudka, {\it {Two-Component
  Dark Matter}},  {\em JHEP} {\bf 10} (2013) 158,
  [\href{http://xxx.lanl.gov/abs/1309.2986}{{\tt 1309.2986}}].

\bibitem{Esch:2014jpa}
S.~Esch, M.~Klasen, and C.~E. Yaguna, {\it {A minimal model for two-component
  dark matter}},  {\em JHEP} {\bf 09} (2014) 108,
  [\href{http://xxx.lanl.gov/abs/1406.0617}{{\tt 1406.0617}}].

\bibitem{Bian:2014cja}
L.~Bian, T.~Li, J.~Shu, and X.-C. Wang, {\it {Two component dark matter with
  multi-Higgs portals}},  {\em JHEP} {\bf 03} (2015) 126,
  [\href{http://xxx.lanl.gov/abs/1412.5443}{{\tt 1412.5443}}].

\bibitem{YaserAyazi:2018lrv}
S.~Yaser~Ayazi and A.~Mohamadnejad, {\it {Scale-Invariant Two Component Dark
  Matter}},  \href{http://xxx.lanl.gov/abs/1808.08706}{{\tt 1808.08706}}.

\bibitem{Ahmed:2017dbb}
A.~Ahmed, M.~Duch, B.~Grzadkowski, and M.~Iglicki, {\it {Multi-Component Dark
  Matter: the vector and fermion case}},  {\em Eur. Phys. J.} {\bf C78} (2018),
  no.~11 905, [\href{http://xxx.lanl.gov/abs/1710.01853}{{\tt 1710.01853}}].

\bibitem{DuttaBanik:2016jzv}
A.~Dutta~Banik, M.~Pandey, D.~Majumdar, and A.~Biswas, {\it {Two component
  WIMP?FImP dark matter model with singlet fermion, scalar and pseudo scalar}},
   {\em Eur. Phys. J.} {\bf C77} (2017), no.~10 657,
  [\href{http://xxx.lanl.gov/abs/1612.08621}{{\tt 1612.08621}}].

\bibitem{DiFranzo:2016uzc}
A.~DiFranzo and G.~Mohlabeng, {\it {Multi-component Dark Matter through a
  Radiative Higgs Portal}},  {\em JHEP} {\bf 01} (2017) 080,
  [\href{http://xxx.lanl.gov/abs/1610.07606}{{\tt 1610.07606}}].

\bibitem{Dienes:2013xff}
K.~R. Dienes, J.~Kumar, and B.~Thomas, {\it {Dynamical Dark Matter and the
  positron excess in light of AMS results}},  {\em Phys. Rev.} {\bf D88}
  (2013), no.~10 103509, [\href{http://xxx.lanl.gov/abs/1306.2959}{{\tt
  1306.2959}}].

\bibitem{Biswas:2015sva}
A.~Biswas, D.~Majumdar, and P.~Roy, {\it {Nonthermal two component dark matter
  model for Fermi-LAT ?-ray excess and 3.55 keV X-ray line}},  {\em JHEP} {\bf
  04} (2015) 065, [\href{http://xxx.lanl.gov/abs/1501.02666}{{\tt
  1501.02666}}].

\bibitem{Herrero-Garcia:2018qnz}
J.~Herrero-Garcia, A.~Scaffidi, M.~White, and A.~G. Williams, {\it {On the
  direct detection of multi-component dark matter: implications of the relic
  abundance}},  {\em JCAP} {\bf 1901} (2019), no.~01 008,
  [\href{http://xxx.lanl.gov/abs/1809.06881}{{\tt 1809.06881}}].

\bibitem{Karam:2016rsz}
A.~Karam and K.~Tamvakis, {\it {Dark Matter from a Classically Scale-Invariant
  $SU(3)_X$}},  {\em Phys. Rev.} {\bf D94} (2016), no.~5 055004,
  [\href{http://xxx.lanl.gov/abs/1607.01001}{{\tt 1607.01001}}].

\bibitem{Bhattacharya:2018cgx}
S.~Bhattacharya, P.~Ghosh, and N.~Sahu, {\it {Multipartite Dark Matter with
  Scalars, Fermions and signatures at LHC}},
  \href{http://xxx.lanl.gov/abs/1809.07474}{{\tt 1809.07474}}.

\bibitem{Bhattacharya:2016ysw}
S.~Bhattacharya, P.~Poulose, and P.~Ghosh, {\it {Multipartite Interacting
  Scalar Dark Matter in the light of updated LUX data}},  {\em JCAP} {\bf 1704}
  (2017), no.~04 043, [\href{http://xxx.lanl.gov/abs/1607.08461}{{\tt
  1607.08461}}].

\bibitem{Dev:2016qeb}
P.~S.~B. Dev, R.~N. Mohapatra, and Y.~Zhang, {\it {Heavy right-handed neutrino
  dark matter in left-right models}},  {\em Mod. Phys. Lett.} {\bf A32} (2017)
  1740007, [\href{http://xxx.lanl.gov/abs/1610.05738}{{\tt 1610.05738}}].

\bibitem{Khlopov:1995pa}
M.~{\relax Yu}. Khlopov, {\it {Physical arguments, favouring multicomponent
  dark matter}},  in {\em {Dark matter in cosmology, clocks and test of
  fundamental laws. Proceedings, 30th Rencontres de Moriond, 15th Moriond
  Workshop, Villars sur Ollon, Switzerland, January 22-29, 1995}},
  pp.~133--138, 1995.

\bibitem{Bhattacharya:2017fid}
S.~Bhattacharya, P.~Ghosh, T.~N. Maity, and T.~S. Ray, {\it {Mitigating Direct
  Detection Bounds in Non-minimal Higgs Portal Scalar Dark Matter Models}},
  {\em JHEP} {\bf 10} (2017) 088,
  [\href{http://xxx.lanl.gov/abs/1706.04699}{{\tt 1706.04699}}].

\bibitem{Huang:2015wts}
W.-C. Huang, Y.-L.~S. Tsai, and T.-C. Yuan, {\it {G2HDM : Gauged Two Higgs
  Doublet Model}},  {\em JHEP} {\bf 04} (2016) 019,
  [\href{http://xxx.lanl.gov/abs/1512.00229}{{\tt 1512.00229}}].

\bibitem{Davoudiasl:2013jma}
H.~Davoudiasl and I.~M. Lewis, {\it {Dark Matter from Hidden Forces}},  {\em
  Phys. Rev.} {\bf D89} (2014), no.~5 055026,
  [\href{http://xxx.lanl.gov/abs/1309.6640}{{\tt 1309.6640}}].

\bibitem{Barman:2018esi}
B.~Barman, S.~Bhattacharya, and M.~Zakeri, {\it {Multipartite Dark Matter in
  $SU(2)_N$ extension of Standard Model and signatures at the LHC}},  {\em
  JCAP} {\bf 1809} (2018), no.~09 023,
  [\href{http://xxx.lanl.gov/abs/1806.01129}{{\tt 1806.01129}}].

\bibitem{Gross:2015cwa}
C.~Gross, O.~Lebedev, and Y.~Mambrini, {\it {Non-Abelian gauge fields as dark
  matter}},  {\em JHEP} {\bf 08} (2015) 158,
  [\href{http://xxx.lanl.gov/abs/1505.07480}{{\tt 1505.07480}}].

\bibitem{Yamanaka:2015tba}
N.~Yamanaka, S.~Fujibayashi, S.~Gongyo, and H.~Iida, {\it {Dark Matter in the
  Nonabelian Hidden Gauge Theory}},  in {\em {2nd Toyama International Workshop
  on Higgs as a Probe of New Physics (HPNP2015) Toyama, Japan, February 11-15,
  2015}}, 2015.
\newblock \href{http://xxx.lanl.gov/abs/1504.08121}{{\tt 1504.08121}}.

\bibitem{Dev:2016xcp}
P.~S. Bhupal~Dev, R.~N. Mohapatra, and Y.~Zhang, {\it {Naturally stable
  right-handed neutrino dark matter}},  {\em JHEP} {\bf 11} (2016) 077,
  [\href{http://xxx.lanl.gov/abs/1608.06266}{{\tt 1608.06266}}].

\bibitem{Blumenthal:1984bp}
G.~R. Blumenthal, S.~M. Faber, J.~R. Primack, and M.~J. Rees, {\it {Formation
  of Galaxies and Large Scale Structure with Cold Dark Matter}},  {\em Nature}
  {\bf 311} (1984) 517--525. [,96(1984)].

\bibitem{Tulin:2017ara}
S.~Tulin and H.-B. Yu, {\it {Dark Matter Self-interactions and Small Scale
  Structure}},  {\em Phys. Rept.} {\bf 730} (2018) 1--57,
  [\href{http://xxx.lanl.gov/abs/1705.02358}{{\tt 1705.02358}}].

\bibitem{Balducci:2018dms}
O.~Balducci, S.~Hofmann, and A.~Kassiteridis, {\it {Small-scale structure from
  charged leptophilia}},  \href{http://xxx.lanl.gov/abs/1812.02182}{{\tt
  1812.02182}}.

\bibitem{Messina:2018fmz}
{\bf XENON} Collaboration, M.~Messina, {\it {Latest results of 1 tonne x year
  Dark Matter Search with XENON1T}},  {\em PoS} {\bf EDSU2018} (2018) 017.

\bibitem{Chakraborty:2012rb}
I.~Chakraborty and A.~Kundu, {\it {Controlling the fine-tuning problem with
  singlet scalar dark matter}},  {\em Phys. Rev.} {\bf D87} (2013), no.~5
  055015, [\href{http://xxx.lanl.gov/abs/1212.0394}{{\tt 1212.0394}}].

\bibitem{Grzadkowski:2009mj}
B.~Grzadkowski and J.~Wudka, {\it {Pragmatic approach to the little hierarchy
  problem: the case for Dark Matter and neutrino physics}},  {\em Phys. Rev.
  Lett.} {\bf 103} (2009) 091802,
  [\href{http://xxx.lanl.gov/abs/0902.0628}{{\tt 0902.0628}}].

\bibitem{Karahan:2014ola}
C.~N. Karahan and B.~Korutlu, {\it {Effects of a Real Singlet Scalar on Veltman
  Condition}},  {\em Phys. Lett.} {\bf B732} (2014) 320--324,
  [\href{http://xxx.lanl.gov/abs/1404.0175}{{\tt 1404.0175}}].

\bibitem{Antipin:2013exa}
O.~Antipin, M.~Mojaza, and F.~Sannino, {\it {Conformal Extensions of the
  Standard Model with Veltman Conditions}},  {\em Phys. Rev.} {\bf D89} (2014),
  no.~8 085015, [\href{http://xxx.lanl.gov/abs/1310.0957}{{\tt 1310.0957}}].

\bibitem{Craig:2013xia}
N.~Craig, C.~Englert, and M.~McCullough, {\it {New Probe of Naturalness}},
  {\em Phys. Rev. Lett.} {\bf 111} (2013), no.~12 121803,
  [\href{http://xxx.lanl.gov/abs/1305.5251}{{\tt 1305.5251}}].

\bibitem{Farina:2013ssa}
M.~Farina, M.~Perelstein, and N.~Rey-Le~Lorier, {\it {Higgs Couplings and
  Naturalness}},  {\em Phys. Rev.} {\bf D90} (2014), no.~1 015014,
  [\href{http://xxx.lanl.gov/abs/1305.6068}{{\tt 1305.6068}}].

\bibitem{Gonderinger:2009jp}
M.~Gonderinger, Y.~Li, H.~Patel, and M.~J. Ramsey-Musolf, {\it {Vacuum
  Stability, Perturbativity, and Scalar Singlet Dark Matter}},  {\em JHEP} {\bf
  01} (2010) 053, [\href{http://xxx.lanl.gov/abs/0910.3167}{{\tt 0910.3167}}].

\bibitem{Drozd:2011aa}
A.~Drozd, B.~Grzadkowski, and J.~Wudka, {\it {Multi-Scalar-Singlet Extension of
  the Standard Model - the Case for Dark Matter and an Invisible Higgs Boson}},
   {\em JHEP} {\bf 04} (2012) 006,
  [\href{http://xxx.lanl.gov/abs/1112.2582}{{\tt 1112.2582}}]. [Erratum:
  JHEP11,130(2014)].

\bibitem{Baek:2012se}
S.~Baek, P.~Ko, W.-I. Park, and E.~Senaha, {\it {Higgs Portal Vector Dark
  Matter : Revisited}},  {\em JHEP} {\bf 05} (2013) 036,
  [\href{http://xxx.lanl.gov/abs/1212.2131}{{\tt 1212.2131}}].

\bibitem{Gabrielli:2013hma}
E.~Gabrielli, M.~Heikinheimo, K.~Kannike, A.~Racioppi, M.~Raidal, and
  C.~Spethmann, {\it {Towards Completing the Standard Model: Vacuum Stability,
  EWSB and Dark Matter}},  {\em Phys. Rev.} {\bf D89} (2014), no.~1 015017,
  [\href{http://xxx.lanl.gov/abs/1309.6632}{{\tt 1309.6632}}].

\bibitem{Hambye:2013sna}
T.~Hambye and A.~Strumia, {\it {Dynamical generation of the weak and Dark
  Matter scale}},  {\em Phys. Rev.} {\bf D88} (2013) 055022,
  [\href{http://xxx.lanl.gov/abs/1306.2329}{{\tt 1306.2329}}].

\bibitem{Noble:2007kk}
A.~Noble and M.~Perelstein, {\it {Higgs self-coupling as a probe of electroweak
  phase transition}},  {\em Phys. Rev.} {\bf D78} (2008) 063518,
  [\href{http://xxx.lanl.gov/abs/0711.3018}{{\tt 0711.3018}}].

\bibitem{Damgaard:2013kva}
P.~H. Damgaard, D.~O'Connell, T.~C. Petersen, and A.~Tranberg, {\it
  {Constraints on New Physics from Baryogenesis and Large Hadron Collider
  Data}},  {\em Phys. Rev. Lett.} {\bf 111} (2013), no.~22 221804,
  [\href{http://xxx.lanl.gov/abs/1305.4362}{{\tt 1305.4362}}].

\bibitem{Profumo:2014opa}
S.~Profumo, M.~J. Ramsey-Musolf, C.~L. Wainwright, and P.~Winslow, {\it
  {Singlet-catalyzed electroweak phase transitions and precision Higgs boson
  studies}},  {\em Phys. Rev.} {\bf D91} (2015), no.~3 035018,
  [\href{http://xxx.lanl.gov/abs/1407.5342}{{\tt 1407.5342}}].

\bibitem{Ambrosi:2017wek}
{\bf DAMPE} Collaboration, G.~Ambrosi {\em et.~al.}, {\it {Direct detection of
  a break in the teraelectronvolt cosmic-ray spectrum of electrons and
  positrons}},  {\em Nature} {\bf 552} (2017) 63--66,
  [\href{http://xxx.lanl.gov/abs/1711.10981}{{\tt 1711.10981}}].

\bibitem{Yuan:2017ysv}
Q.~Yuan {\em et.~al.}, {\it {Interpretations of the DAMPE electron data}},
  \href{http://xxx.lanl.gov/abs/1711.10989}{{\tt 1711.10989}}.

\bibitem{Fan:2017sor}
Y.-Z. Fan, W.-C. Huang, M.~Spinrath, Y.-L.~S. Tsai, and Q.~Yuan, {\it {A model
  explaining neutrino masses and the DAMPE cosmic ray electron excess}},  {\em
  Phys. Lett.} {\bf B781} (2018) 83--87,
  [\href{http://xxx.lanl.gov/abs/1711.10995}{{\tt 1711.10995}}].

\bibitem{Duan:2017pkq}
G.~H. Duan, L.~Feng, F.~Wang, L.~Wu, J.~M. Yang, and R.~Zheng, {\it {Simplified
  TeV leptophilic dark matter in light of DAMPE data}},  {\em JHEP} {\bf 02}
  (2018) 107, [\href{http://xxx.lanl.gov/abs/1711.11012}{{\tt 1711.11012}}].

\bibitem{Gu:2017gle}
P.-H. Gu and X.-G. He, {\it {Electrophilic dark matter with dark photon: from
  DAMPE to direct detection}},  {\em Phys. Lett.} {\bf B778} (2018) 292--295,
  [\href{http://xxx.lanl.gov/abs/1711.11000}{{\tt 1711.11000}}].

\bibitem{Cao:2017ydw}
J.~Cao, L.~Feng, X.~Guo, L.~Shang, F.~Wang, and P.~Wu, {\it {Scalar dark matter
  interpretation of the DAMPE data with U(1) gauge interactions}},  {\em Phys.
  Rev.} {\bf D97} (2018), no.~9 095011,
  [\href{http://xxx.lanl.gov/abs/1711.11452}{{\tt 1711.11452}}].

\bibitem{Liu:2017rgs}
X.~Liu and Z.~Liu, {\it {TeV dark matter and the DAMPE electron excess}},  {\em
  Phys. Rev.} {\bf D98} (2018), no.~3 035025,
  [\href{http://xxx.lanl.gov/abs/1711.11579}{{\tt 1711.11579}}].

\bibitem{Tang:2017lfb}
Y.-L. Tang, L.~Wu, M.~Zhang, and R.~Zheng, {\it {Lepton-portal Dark Matter in
  Hidden Valley model and the DAMPE recent results}},  {\em Sci. China Phys.
  Mech. Astron.} {\bf 61} (2018), no.~10 101003,
  [\href{http://xxx.lanl.gov/abs/1711.11058}{{\tt 1711.11058}}].

\bibitem{Chao:2017yjg}
W.~Chao and Q.~Yuan, {\it {The electron-flavored Z'-portal dark matter and the
  DAMPE cosmic ray excess}},  \href{http://xxx.lanl.gov/abs/1711.11182}{{\tt
  1711.11182}}.

\bibitem{Gu:2017bdw}
P.-H. Gu, {\it {Radiative Dirac neutrino mass, DAMPE dark matter and
  leptogenesis}},  \href{http://xxx.lanl.gov/abs/1711.11333}{{\tt 1711.11333}}.

\bibitem{Duan:2017qwj}
G.~H. Duan, X.-G. He, L.~Wu, and J.~M. Yang, {\it {Leptophilic dark matter in
  gauged $U(1)_{L{_e}-L_{\mu }}$ model in light of DAMPE cosmic ray ${e{^+}} +
  {e{^-}}$ excess}},  {\em Eur. Phys. J.} {\bf C78} (2018), no.~4 323,
  [\href{http://xxx.lanl.gov/abs/1711.11563}{{\tt 1711.11563}}].

\bibitem{Jin:2017qcv}
H.-B. Jin, B.~Yue, X.~Zhang, and X.~Chen, {\it {Dark matter explanation of the
  cosmic ray $e^{+} e^{-}$ spectrum excess and peak feature observed by the
  DAMPE experiment}},  {\em Phys. Rev.} {\bf D98} (2018), no.~12 123008,
  [\href{http://xxx.lanl.gov/abs/1712.00362}{{\tt 1712.00362}}].

\bibitem{Niu:2017hqe}
J.-S. Niu, T.~Li, R.~Ding, B.~Zhu, H.-F. Xue, and Y.~Wang, {\it {Bayesian
  analysis of the break in $DAMPE$ lepton spectra}},  {\em Phys. Rev.} {\bf
  D97} (2018), no.~8 083012, [\href{http://xxx.lanl.gov/abs/1712.00372}{{\tt
  1712.00372}}].

\bibitem{Li:2017tmd}
T.~Li, N.~Okada, and Q.~Shafi, {\it {Scalar dark matter, Type II Seesaw and the
  DAMPE cosmic ray $e^+ + e^-$ excess}},  {\em Phys. Lett.} {\bf B779} (2018)
  130--135, [\href{http://xxx.lanl.gov/abs/1712.00869}{{\tt 1712.00869}}].

\bibitem{Gu:2017lir}
P.-H. Gu, {\it {Quasi-degenerate dark matter for DAMPE excess and 3.5 keV
  line}},  {\em Sci. China Phys. Mech. Astron.} {\bf 61} (2018), no.~10 101005,
  [\href{http://xxx.lanl.gov/abs/1712.00922}{{\tt 1712.00922}}].

\bibitem{Nomura:2017ohi}
T.~Nomura and H.~Okada, {\it {Radiative seesaw models linking to dark matter
  candidates inspired by the DAMPE excess}},  {\em Phys. Dark Univ.} {\bf 21}
  (2018) 90--95, [\href{http://xxx.lanl.gov/abs/1712.00941}{{\tt 1712.00941}}].

\bibitem{Ghorbani:2017cey}
K.~Ghorbani and P.~H. Ghorbani, {\it {DAMPE electron-positron excess in
  leptophilic Z? model}},  {\em JHEP} {\bf 05} (2018) 125,
  [\href{http://xxx.lanl.gov/abs/1712.01239}{{\tt 1712.01239}}].

\bibitem{Yang:2017cjm}
F.~Yang, M.~Su, and Y.~Zhao, {\it {Dark Matter Annihilation from Nearby
  Ultra-compact Micro Halos to Explain the Tentative Excess at ~1.4 TeV in
  DAMPE data}},  \href{http://xxx.lanl.gov/abs/1712.01724}{{\tt 1712.01724}}.

\bibitem{Ding:2017jdr}
R.~Ding, Z.-L. Han, L.~Feng, and B.~Zhu, {\it {Confronting the DAMPE Excess
  with the Scotogenic Type-II Seesaw Model}},  {\em Chin. Phys.} {\bf C42}
  (2018), no.~8 083104, [\href{http://xxx.lanl.gov/abs/1712.02021}{{\tt
  1712.02021}}].

\bibitem{Okada:2017pgr}
N.~Okada and O.~Seto, {\it {DAMPE excess from decaying right-handed neutrino
  dark matter}},  {\em Mod. Phys. Lett.} {\bf A33} (2018), no.~27 1850157,
  [\href{http://xxx.lanl.gov/abs/1712.03652}{{\tt 1712.03652}}].

\bibitem{Yao:2018ewe}
Y.-h. Yao, C.~Jin, and X.-c. Chang, {\it {Test of the 1.4 TeV DAMPE electron
  excess with preliminary H.E.S.S. measurement}},  {\em Nucl. Phys.} {\bf B934}
  (2018) 396--407.

\bibitem{Beck:2018hau}
G.~Beck and S.~Colafrancesco, {\it {Dark matter gets DAMPE}},  in {\em {61st
  Annual Conference of the South African Institute of Physics (SAIP2016)
  Johannesburg, South Africa, July 4-8, 2016}}, 2018.
\newblock \href{http://xxx.lanl.gov/abs/1810.07176}{{\tt 1810.07176}}.

\bibitem{Wang:2018pcc}
B.~Wang, X.~Bi, S.~Lin, and P.~Yin, {\it {Explanations of the DAMPE high energy
  electron/positron spectrum in the dark matter annihilation and pulsar
  scenarios}},  {\em Sci. China Phys. Mech. Astron.} {\bf 61} (2018), no.~10
  101004.

\bibitem{Cao:2017sju}
J.~Cao, L.~Feng, X.~Guo, L.~Shang, F.~Wang, P.~Wu, and L.~Zu, {\it {Explaining
  the DAMPE data with scalar dark matter and gauged $U(1)_{L_e-L_\mu }$
  interaction}},  {\em Eur. Phys. J.} {\bf C78} (2018), no.~3 198,
  [\href{http://xxx.lanl.gov/abs/1712.01244}{{\tt 1712.01244}}].

\bibitem{Cao:2017rjr}
J.~Cao, X.~Guo, L.~Shang, F.~Wang, P.~Wu, and L.~Zu, {\it {Scalar dark matter
  explanation of the DAMPE data in the minimal Left-Right symmetric model}},
  {\em Phys. Rev.} {\bf D97} (2018), no.~6 063016,
  [\href{http://xxx.lanl.gov/abs/1712.05351}{{\tt 1712.05351}}].

\bibitem{Weinberg:1979sa}
S.~Weinberg, {\it {Baryon and Lepton Nonconserving Processes}},  {\em Phys.
  Rev. Lett.} {\bf 43} (1979) 1566--1570.

\bibitem{Mohapatra:1979ia}
R.~N. Mohapatra and G.~Senjanovic, {\it {Neutrino Mass and Spontaneous Parity
  Nonconservation}},  {\em Phys. Rev. Lett.} {\bf 44} (1980) 912. [,231(1979)].

\bibitem{Kannike:2012pe}
K.~Kannike, {\it {Vacuum Stability Conditions From Copositivity Criteria}},
  {\em Eur. Phys. J.} {\bf C72} (2012) 2093,
  [\href{http://xxx.lanl.gov/abs/1205.3781}{{\tt 1205.3781}}].

\bibitem{Schael:2013ita}
{\bf ALEPH, DELPHI, L3, OPAL, LEP Electroweak} Collaboration, S.~Schael {\em
  et.~al.}, {\it {Electroweak Measurements in Electron-Positron Collisions at
  W-Boson-Pair Energies at LEP}},  {\em Phys. Rept.} {\bf 532} (2013) 119--244,
  [\href{http://xxx.lanl.gov/abs/1302.3415}{{\tt 1302.3415}}].

\bibitem{Ade:2015xua}
{\bf Planck} Collaboration, P.~A.~R. Ade {\em et.~al.}, {\it {Planck 2015
  results. XIII. Cosmological parameters}},  {\em Astron. Astrophys.} {\bf 594}
  (2016) A13, [\href{http://xxx.lanl.gov/abs/1502.01589}{{\tt 1502.01589}}].

\bibitem{Springel:2005nw}
V.~Springel {\em et.~al.}, {\it {Simulating the joint evolution of quasars,
  galaxies and their large-scale distribution}},  {\em Nature} {\bf 435} (2005)
  629--636, [\href{http://xxx.lanl.gov/abs/astro-ph/0504097}{{\tt
  astro-ph/0504097}}].

\bibitem{Berlin:2014tja}
A.~Berlin, D.~Hooper, and S.~D. McDermott, {\it {Simplified Dark Matter Models
  for the Galactic Center Gamma-Ray Excess}},  {\em Phys. Rev.} {\bf D89}
  (2014), no.~11 115022, [\href{http://xxx.lanl.gov/abs/1404.0022}{{\tt
  1404.0022}}].

\bibitem{Ko:2014gha}
P.~Ko, W.-I. Park, and Y.~Tang, {\it {Higgs portal vector dark matter for
  $\mathinner{\mathrm{GeV}}$ scale $\gamma$-ray excess from galactic center}},
  {\em JCAP} {\bf 1409} (2014) 013,
  [\href{http://xxx.lanl.gov/abs/1404.5257}{{\tt 1404.5257}}].

\bibitem{Barger:2007im}
V.~Barger, P.~Langacker, M.~McCaskey, M.~J. Ramsey-Musolf, and G.~Shaughnessy,
  {\it {LHC Phenomenology of an Extended Standard Model with a Real Scalar
  Singlet}},  {\em Phys. Rev.} {\bf D77} (2008) 035005,
  [\href{http://xxx.lanl.gov/abs/0706.4311}{{\tt 0706.4311}}].

\bibitem{Djouadi:2012zc}
A.~Djouadi, A.~Falkowski, Y.~Mambrini, and J.~Quevillon, {\it {Direct Detection
  of Higgs-Portal Dark Matter at the LHC}},  {\em Eur. Phys. J.} {\bf C73}
  (2013), no.~6 2455, [\href{http://xxx.lanl.gov/abs/1205.3169}{{\tt
  1205.3169}}].

\bibitem{Casas:2017jjg}
J.~A. Casas, D.~G. Cerdeño, J.~M. Moreno, and J.~Quilis, {\it {Reopening the
  Higgs portal for single scalar dark matter}},  {\em JHEP} {\bf 05} (2017)
  036, [\href{http://xxx.lanl.gov/abs/1701.08134}{{\tt 1701.08134}}].

\bibitem{Ackermann:2013yva}
{\bf Fermi-LAT} Collaboration, M.~Ackermann {\em et.~al.}, {\it {Dark matter
  constraints from observations of 25 Milky Way satellite galaxies with the
  Fermi Large Area Telescope}},  {\em Phys. Rev.} {\bf D89} (2014) 042001,
  [\href{http://xxx.lanl.gov/abs/1310.0828}{{\tt 1310.0828}}].

\bibitem{Fowlie:2017fya}
A.~Fowlie, {\it {DAMPE squib? Significance of the 1.4 TeV DAMPE excess}},  {\em
  Phys. Lett.} {\bf B780} (2018) 181--184,
  [\href{http://xxx.lanl.gov/abs/1712.05089}{{\tt 1712.05089}}].

\bibitem{Huang:2017egk}
X.-J. Huang, Y.-L. Wu, W.-H. Zhang, and Y.-F. Zhou, {\it {Origins of sharp
  cosmic-ray electron structures and the DAMPE excess}},  {\em Phys. Rev.} {\bf
  D97} (2018), no.~9 091701, [\href{http://xxx.lanl.gov/abs/1712.00005}{{\tt
  1712.00005}}].

\bibitem{Niu:2017lts}
J.-S. Niu, T.~Li, and F.-Z. Xu, {\it {A Simple and Natural Interpretations of
  the DAMPE Cosmic Ray Electron/Positron Spectrum within Two Sigma
  Deviations}},  \href{http://xxx.lanl.gov/abs/1712.09586}{{\tt 1712.09586}}.

\bibitem{Ge:2017tkd}
S.-F. Ge, H.-J. He, and Y.-C. Wang, {\it {Flavor Structure of the Cosmic-Ray
  Electron/Positron Excesses at DAMPE}},  {\em Phys. Lett.} {\bf B781} (2018)
  88--94, [\href{http://xxx.lanl.gov/abs/1712.02744}{{\tt 1712.02744}}].

\bibitem{Chao:2017emq}
W.~Chao, H.-K. Guo, H.-L. Li, and J.~Shu, {\it {Electron Flavored Dark
  Matter}},  {\em Phys. Lett.} {\bf B782} (2018) 517--522,
  [\href{http://xxx.lanl.gov/abs/1712.00037}{{\tt 1712.00037}}].

\bibitem{Nomura:2018jkd}
T.~Nomura, H.~Okada, and P.~Wu, {\it {A radiative neutrino mass model in light
  of DAMPE excess with hidden gauged $U(1)$ symmetry}},  {\em JCAP} {\bf 1805}
  (2018), no.~05 053, [\href{http://xxx.lanl.gov/abs/1801.04729}{{\tt
  1801.04729}}].

\bibitem{Zhao:2017nrt}
Y.~Zhao, K.~Fang, M.~Su, and M.~C. Miller, {\it {A Strong Test of the Dark
  Matter Origin of the 1.4 TeV DAMPE Signal Using IceCube Neutrinos}},
  \href{http://xxx.lanl.gov/abs/1712.03210}{{\tt 1712.03210}}.

\bibitem{Giusarma:2016phn}
E.~Giusarma, M.~Gerbino, O.~Mena, S.~Vagnozzi, S.~Ho, and K.~Freese, {\it
  {Improvement of cosmological neutrino mass bounds}},  {\em Phys. Rev.} {\bf
  D94} (2016), no.~8 083522, [\href{http://xxx.lanl.gov/abs/1605.04320}{{\tt
  1605.04320}}].

\bibitem{Giusarma:2018jei}
E.~Giusarma, S.~Vagnozzi, S.~Ho, S.~Ferraro, K.~Freese, R.~Kamen-Rubio, and
  K.-B. Luk, {\it {Scale-dependent galaxy bias, CMB lensing-galaxy
  cross-correlation, and neutrino masses}},  {\em Phys. Rev.} {\bf D98} (2018),
  no.~12 123526, [\href{http://xxx.lanl.gov/abs/1802.08694}{{\tt 1802.08694}}].

\bibitem{Shtabovenko:2016sxi}
V.~Shtabovenko, R.~Mertig, and F.~Orellana, {\it {New Developments in FeynCalc
  9.0}},  {\em Comput. Phys. Commun.} {\bf 207} (2016) 432--444,
  [\href{http://xxx.lanl.gov/abs/1601.01167}{{\tt 1601.01167}}].

\end{thebibliography}\endgroup
\bibliographystyle{JHEP}

\end{document}